%% file: fse-25.tex
\documentclass[acmsmall,screen]{acmart}
\AtBeginDocument{%
  \providecommand\BibTeX{{%
    \normalfont B\kern-0.5em{\scshape i\kern-0.25em b}\kern-0.8em\TeX}}}

\usepackage{listings}
\usepackage{colortbl, xcolor}
\usepackage{xcolor}
\usepackage{diagbox}
\usepackage{makecell}
\definecolor{pblue}{rgb}{0.13,0.13,1}
\definecolor{pgreen}{rgb}{0,0.5,0}
\definecolor{pred}{rgb}{0.9,0,0}
\definecolor{pgrey}{rgb}{0.46,0.45,0.48}
\definecolor{applegreen}{rgb}{0, 0.5, 0.0}
\usepackage[flushleft]{threeparttable}
\usepackage{tabularx}
\usepackage{subcaption}
\usepackage{makecell}
\usepackage{tcolorbox}
\usepackage{multirow}
\usepackage{array}
\usepackage{enumitem}
\usepackage{graphicx}
\usepackage{soul}
\usepackage{tabularx}
\usepackage{makecell}
\usepackage{mdframed}
\usepackage{xspace}
\usepackage{adjustbox}

\definecolor{codeblue}{RGB}{20,76,134}
\definecolor{codegreysh}{RGB}{114,136,223}
\definecolor{code}{RGB}{51,51,255}
\lstdefinestyle{listingstyle}{
    language=Java,
    basicstyle=\ttfamily\scriptsize,
    keywordstyle=\bf\ttfamily\color{codeblue},
    stringstyle=\color{codegreysh},
    moredelim=[l][\bf\ttfamily\color{red}]{///},
    moredelim=[l][\bf\ttfamily\color{orange}]{//,},
    moredelim=[s][\bf\ttfamily\color{code}]{/**}{**/}
}

\newcommand{\doc}{Javadoc\xspace}

\setcopyright{acmlicensed}
\copyrightyear{2025}
\acmYear{2025}
\acmDOI{XXXXXXX.XXXXXXX}

\acmISBN{978-1-4503-XXXX-X/18/06}

\begin{document}

%%
%% The "title" command has an optional parameter,
%% allowing the author to define a "short title" to be used in page headers.
\title{Doc2OracLL: Investigating the Impact of Documentation on LLM-based Test Oracle Generation}

\author{Soneya Binta Hossain}
\orcid{0000-0002-7282-061X}
\affiliation{%
  \institution{University of Virginia}
  \city{Charlottesville}
  \country{USA}
}
\email{sh7hv@virginia.edu}

\author{Raygan Taylor}
\affiliation{%
  \institution{Dillard University}
  \city{New Orleans}
  \country{USA}
}
\email{raygan.taylor@dillard.edu}

\author{Matthew Dwyer}
\affiliation{%
  \institution{University of Virginia}
  \city{Charlottesville}
  \country{USA}
}
\email{md3cn@virginia.edu}

%%
%% By default, the full list of authors will be used in the page
%% headers. Often, this list is too long, and will overlap
%% other information printed in the page headers. This command allows
%% the author to define a more concise list
%% of authors' names for this purpose.

\input{macros}
\setlist[itemize]{leftmargin=20pt}

%%
%% The abstract is a short summary of the work to be presented in the
%% article.

\begin{abstract}

Code documentation is a critical artifact of software development, bridging human understanding and machine-readable code. Beyond aiding developers in code comprehension and maintenance, documentation also plays a critical role in \textit{automating various software engineering tasks, such as test oracle generation (TOG)}. In Java, \doc comments offer structured, natural language documentation embedded directly within the source code, typically describing functionality, usage, parameters, return values, and exceptional behavior. \mrnew{While prior research has explored the use of \doc comments in TOG alongside other information, such as the method under test, their potential as a stand-alone input source, the most relevant \doc components, and guidelines for writing effective \doc comments for automating TOG remain less explored.}.

In this study, we investigate the impact of Javadoc comments on TOG through a comprehensive analysis. We begin by fine-tuning 10 large language models using three different prompt pairs to assess the role of Javadoc comments alongside other contextual information. Next, we systematically analyze the impact of different Javadoc comment's \textbf{components} on TOG. To evaluate the generalizability of Javadoc comments from various sources, we also generate them using the GPT-3.5 model. We perform a thorough bug detection study using \textit{Defects4J} dataset to understand their role in real-world bug detection. \mr{Our results show that incorporating Javadoc comments improves the accuracy of test oracles in most cases, aligning closely with ground truth. We find that Javadoc comments \textit{alone} can achieve comparable or even better performance when using 
the implementation of MUT. Additionally, we identify that the \textbf{description} and the \textbf{return tag} are the most valuable components for TOG. Finally, our approach, when using only Javadoc comments, detects between 19\% and 94\% more real-world bugs in Defects4J than prior methods, establishing a new state-of-the-art}. \mrnew{To further guide developers in writing effective documentation, we conduct a detailed qualitative study on when Javadoc comments are helpful or harmful for TOG.}

%In this study, we dive deep into investigating the impact of \doc comments on TOG. We start by \mr{fine-tuning} 10 large language models with three different prompt pairs designed to investigate the impact of \doc comments when using with other contextual information. We conduct a systematic analysis to assess the impact of different \doc components on TOG. For investigating the generalizability of the \doc comments from various sources, we also generate \doc comments using GPT-3.5 model. Finally, we perform a thorough bug detection study using Defects4J to understand the role of \doc comments in real-world bug detection. Our results show that, in most cases, incorporating \doc comments improves the accuracy of test oracles, aligning closely with ground truth. We find that \textit{\doc comments alone can nearly match the performance achieved when using both \doc comments and MUT code. We also find that the \textbf{description} and the \textbf{return tag} of the \doc comment are most valuable in TOG. Finally, when using just \doc comments our method \mr{detected} between 19\% and 94\% more real-world bugs in Defects4J than prior methods, \mr{establishing itself as the new state-of-the-art.}} \mr{We also conduct a detailed qualitative study to inform developers about when \doc comments are helpful or harmful for TOG.}
\end{abstract}




\maketitle

\section{Introduction}

Testing is a critical step in the software development process.  It involves solving two challenges: (1) \textit{selecting test inputs} to adequately exercise program behavior,
and (2) judging whether the program produces the desired result for those inputs -- the \textit{test oracle} problem~\cite{barr2014oracle}. The past decade has witnessed enormous advances in automating support for the first of these~\cite{pacheco2007randoop,fraser2014evosuite,lemieux2018fairfuzz,bohme2019coverage} and such techniques are now regularly used in practice.

Providing effective
automated support for the oracle problem has been less successful,
though in recent years, researchers
have made some progress using machine learning (ML) techniques~\cite{dinella2022toga,schafer2023empirical,hossain2024togll,endres2024can}.
All of these techniques either require~\cite{dinella2022toga,schafer2023empirical} or allow~\cite{hossain2023neural,endres2024can} for the implementation
of the method under test (MUT) to be included in the ML model's input
to generate a test oracle.
Recent research~\cite{liu2023towards} has pointed out that this
may negatively impact the quality of the test oracle. For example, if the implementation contains a bug, the model may encode that buggy behavior into the oracle, resulting in an inappropriately passing test case. More generally, this approach tends to generate oracles that are primarily useful in a regression context.

One way to address this weakness is to leverage good documentation for the MUT.
Conceptually, good method-level comments
provide a correct and complete, yet abstract, description of intended behavior that is decoupled from implementation detail~\cite{aghajani2019software}.
Such code comments are widely acknowledged by practitioners
as useful for ``development and testing'' tasks~\cite{aghajani2020software}.
Moreover practitioners find that method-level comments
are the most valuable of all forms of software documentation and,
for such comments, information about
``functionality'', ``usage'', and ``input \& output'' is most
important~\cite{hu2022practitioners}.
We observe that this type of information aligns with the needs of software testers, who aim to select test inputs based on appropriate method usage and define test oracles \mr{to} assess whether outputs match the intended functionality.

\begin{figure}[t]
    \small\centering
\includegraphics[width=.9\columnwidth]{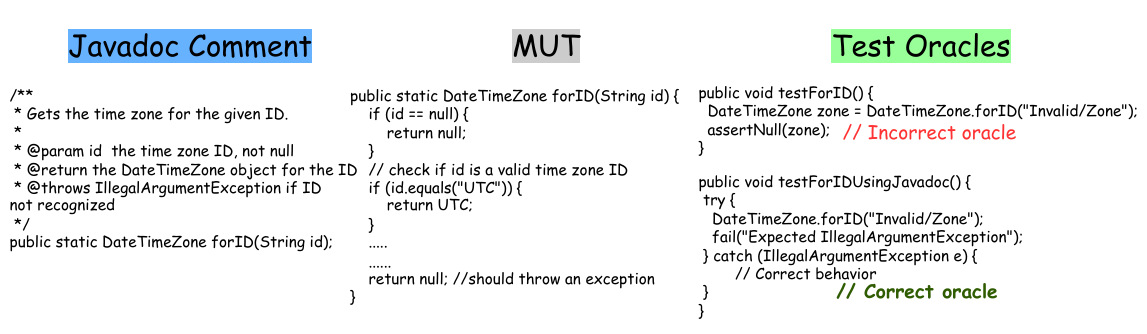}
	\caption{\mr{Incorrect oracle generated from buggy MUT and correct oracle generated from \doc comments. }}
	\label{fig:example}
\end{figure}

Figure~\ref{fig:example} shows an example from the Joda-Time library that illustrates the risk of using the MUT and the benefit of using \doc instead. The MUT returns \texttt{null} when the time zone id is not recognized, where as based on the \doc comments, it should throw an \texttt{IllegalArgumentException}. When using the MUT code to generate test oracles, it generates oracles based on the faulty MUT implementation and is thus unable to detect the bug. When using the \doc comments only, the generated oracle expects the correct behavior that an \texttt{IllegalArgumentException} should be thrown. 

In this paper, we study the value of method-level documentation
in test oracle generation (TOG) using large-language models (LLM). 
We use \doc comments as method-level documentation, which are specifically designed to document Java classes, methods, and interfaces. These comments are written in a structured format using natural language, documenting detailed information about the expected behavior of method, its parameters and return value, and exceptional behavior. Their structured format, combined with rich contextual information, makes \doc comments highly suitable for LLMs. The first sentence of each \doc comment consist of a concise but complete description of the behavior of the MUT -- we refer to this as the description. 
After that, \doc comment may include @param tag with parameter description, @return tag with return value details, @throws/@exception tag detailing exceptional behavior, and some other tags, such as, the @see tag which introduces cross-references to other methods~\cite{javadoc}.

\mr{We build on recent work~\cite{hossain2024togll} that explored a range of prompt formats to fine-tune LLMs for test oracles generation task.}
Unlike that work, our focus here is on the role of the \doc in test oracle generation.
While \cite{hossain2024togll} did evaluate the differential benefit of including \doc comments
in prompts, that work did not account for the fact that 60\% of the dataset used in
their evaluation had no \doc.  This caused their study to substantially underestimate
the benefit of leveraging \doc in test oracle generation. Furthermore, their work did not explore the impact of various components of \doc comments on TOG, nor did they examine the role of \doc comments in bug detection.

We address this by developing several dataset variants in which every MUT
has \doc comments.  The first dataset variant selects more than 55-thousand samples
for which developer written \doc comments are available.
It has been well-documented that the quality of method-level documentation varies significantly in practice~\cite{wang2023suboptimal}, but short of human subject studies, e.g.,~\cite{aghajani2019software,aghajani2020software,hu2022practitioners}, 
determining documentation quality is very challenging especially in light of
recent results demonstrating the unreliability of natural language similarity metrics for this problem~\cite{roy2021reassessing}.
Consequently, we generate a second dataset variant,
inspired by research on code summarization~\cite{mcburney2015automatic,leclair2019neural},
that generates \doc comment that summarizes the MUT using a generative model.
Studies using these two dataset variants show that incorporating \doc comments
for the test oracle generation task leads to a 10-20\% improvement in the ability
to produce \mr{ground-truth test oracles, both assertion and exception.}

\doc is a rich documentation format with many different tag types as shown in Figure~\ref{fig:example} and LLMs produce
higher-quality responses when prompted with the most relevant information for a task.
This led us to conduct an ablation study on the structure of \doc comments in which we
systematically dropped the description and tags to understand their differential value
in the TOG task.  We find clear evidence that the description and @return tag provide
the greatest value, which suggests that when LLM token limits come into play in prompting
prioritizing these \doc elements in a prompt is critical for high-quality TOG.

Finally, while the bulk of our study is conducted on a large and broadly representative
dataset, we go further to understand how fine-tuned LLMs perform TOG generation on 
unseen datasets that harbor real-world bugs.  In a study on Defects4J~\cite{just2014defects4j},
we demonstrate that our methods generate test oracles that are able to detect up to
44\% more bugs than prior approaches~\cite{dinella2022toga,endres2024can}, while
not leveraging the MUT implementation.  When prior approaches do not use the MUT their
performance degrades and our method can detect 94\% more bugs.

\mr{\textbf{The primary contributions of this paper are:}  
 (1) \textbf{Empirical demonstration} across a wide range of fine-tuned LLMs and two distinct documentation sources, showing that method-level documentation improves TOG’s ability to match ground truth oracles.  
  (2) \textbf{Ablation study} on \doc comment structure, identifying \textbf{descriptions} and \textbf{\texttt{@return}} tags as the most valuable components for TOG.  
   (3)  \textbf{Bug detection analysis}, showing that fine-tuned models leveraging \doc comments can generate test oracles that detect significantly more bugs than prior TOG approaches relying on MUT code.}  
  \mrnew{(4) \textbf{Qualitative analysis} providing guidelines on writing effective documentation to enable the generation of stronger test oracles.}

\mrnew{\noindent \textbf{Collectively, these contributions provide compelling evidence that TOG can be performed using software artifacts that exclude the method implementation, reducing a major source of bias in prior work.}}

\section{Approach}

\begin{figure}[t]
    \small\centering
\includegraphics[width=.8\columnwidth]{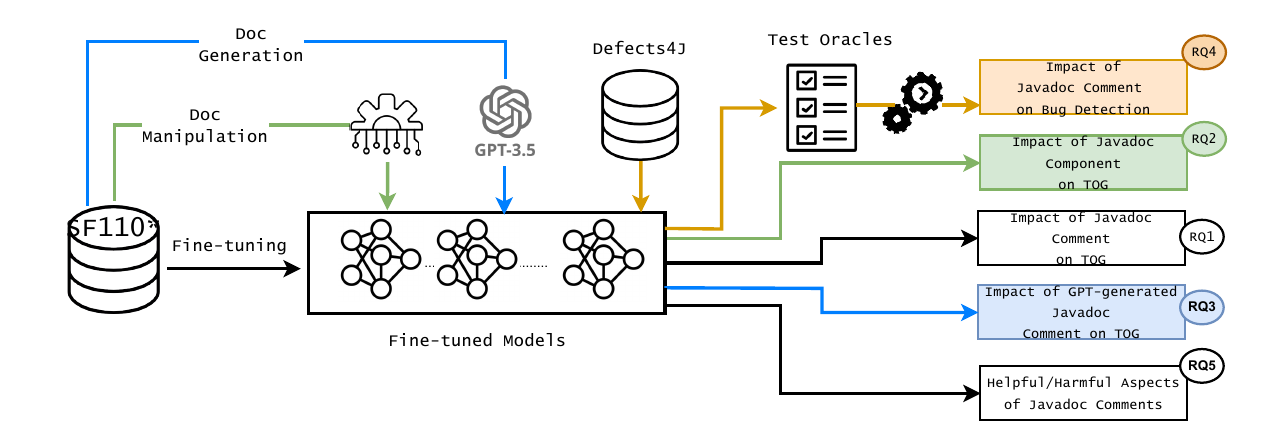}
	\caption{Overview of our approach \mr{Doc2OracLL.}}
	\label{fig:approach}
\end{figure}

In Figure \ref{fig:approach}, we show an overview of our approach. We first fine-tune several large language models, as discussed in Section~\ref{codemodels}, using three pairs of prompts described in Section~\ref{prompts}. To fine-tune the LLMs on these prompts, we employ the dataset detailed in Section~\ref{training-data}. We generate various version of the test dataset by manipulating developer written \doc comments and by generating alternative \doc using GPT-3.5 model. We also perform unseen inference in the context of a real-bug detection study to investigate the impact of \doc comments. \mr{In this section, we provide an overview of the dataset, the prompts, the LLMs, and the metrics used to evaluate our method and compare it with other approaches.}

\subsection{Supervised Fine-tuning}\label{fine-tuning} This section discusses the dataset, prompts, and LLMs that are fine-tuned to investigate the impact of \doc comments on \mr{the} test oracle generation (TOG) task.

\subsubsection{SF110* Dataset}\label{training-data} To construct our dataset, we utilized the SF110 benchmark~\cite{fraser2014evosuite}, which comprises 110 real-world open-source Java projects from various domains, \mr{all} sourced from the SourceForge repository. The benchmark contains a total of 23,000 Java classes and is widely used for unit test generation~\cite{hossain2024togll,siddiq2024using}. The dataset consists of tuples $((p_i, m_i, d_i), o_i)$, where $p_i$ is the test prefix, $m_i$ is the MUT, $d_i$ is the \doc comments (if available), and $o_i$ is the ground truth test oracle, either an assertion or an exception. In total, the dataset includes 140,514 samples, of which 55,575 (40\%) contain \doc comments. These samples with \doc comments are used for fine-tuning the LLMs and have been split into three distinct subsets: 90\% for training, 5\% for validation, and 5\% for testing.

%upto this 
\subsubsection{\mr{Prompts}}\label{prompts} The primary goal of our study is to investigate the impact of \doc comments on automated \mr{TOG}. To ensure \mr{our experiments are \textit{well-controlled} and to \textit{isolate}} the effect of \doc comments, we designed three pairs of prompts. In each prompt pair, the only \mr{\textit{difference}} between the two prompts is \mr{\textit{whether \doc comments are present or absent}.}

%The primary goal of our study is to thoroughly investigate the impact of \doc comments on automated test oracle generation. To ensure the experiments are well-controlled and isolate the effect of \doc comments, we designed three pairs of prompts. In each prompt pair, the only \textit{difference} between the two prompts is \textit{the presence or absence of \doc comments}. 

\begin{itemize}

\item \mr{\textbf{Pair 1 ($\mathcal{P}{1}$, $\mathcal{P}{2}$)}: In this prompt pair, both prompts share the same minimal input—namely, the test prefix. However, prompt $\mathcal{P}_{2}$ also includes the \doc comments.} By comparing the accuracy of these two prompts, we can directly assess the impact of \doc comments when only minimal context—the test prefix—is provided.

\item \mr{\textbf{Pair 2 ($\mathcal{P}{3}$, $\mathcal{P}{4}$)}}: \mr{In this prompt pair, both prompts include two key pieces of information—the test prefix and the MUT signature—while prompt $\mathcal{P}_{4}$ adds the \doc comments as extra context.} By comparing the accuracy of these two prompts, we can gauge the influence of \doc comments when both the test prefix and MUT signature are provided. \mr{Compared to Pair 1, this pair further elucidates how additional contextual information (the MUT signature) affects the role and impact of \doc comments.}

\item \mr{\textbf{Pair 3 ($\mathcal{P}_{5}$, $\mathcal{P}_{6}$)}}: In this prompt pair, both prompts include the test prefix and the full MUT (Method Under Test) code. However, prompt $\mathcal{P}_{6}$ \mr{also} provides additional \doc comments. By comparing the performance of these two prompts, we can analyze the contribution of \doc comments when the model already has access to the complete MUT code. This comparison \mr{clarifies} the added value of \doc comments when substantial contextual information is already available \mr{from} both the test prefix and the \mr{MUT code}, further refining our insights from \mr{prompt pairs} 1 and 2.

\end{itemize}
\smallskip
\subsubsection{Code Models}\label{codemodels} \mr{Decoder-only models are widely used for code generation~\cite{hou2023large}, and smaller models offer several advantages over larger ones. For instance, fine-tuning smaller models generally requires fewer GPU resources, making them more cost-effective and reducing their environmental impact~\cite{bender2021dangers}. In this study, we selected four smaller models (fewer than 1 billion parameters), four models ranging from 1 to 2.7 billion parameters, and two models with 7 billion parameters. All of these models are decoder-only, pre-trained on multiple programming languages, and publicly available on Hugging Face\cite{huggingface2024}. They also come with sufficient documentation for fine-tuning, which has led to their adoption across a variety of tasks throughout the software development lifecycle (SDLC)\cite{10.1145/3660773, hou2023large}}.

\smallskip

\noindent\textbf{CodeGPT~\cite{lu2021codexglue}}:
In our study, we used the \mr{\textit{microsoft/CodeGPT-small-java}} model, a GPT-style language model with 110 million parameters. \mr{This model was trained and tested on a dataset of Java programs and is pre-trained for the code completion task, which aligns well with our designed prompts.}

\smallskip

\noindent\textbf{CodeParrot~\cite{codeparrot}}: \mr{We fine-tuned the 110-million-parameter version of CodeParrot, which was originally pre-trained for code generation on nine different programming languages, including \texttt{Java}, \texttt{C++}, and \texttt{Python}. Specifically, we employed the \textit{codeparrot/codeparrot-small-multi} model from Hugging Face. Given its strong generalization on unseen data and its ability to outperform larger models \cite{hossain2024togll}, we selected this model for the TOG task.}
\smallskip

\noindent\textbf{CodeGen~\cite{nijkamp2023codegen}}: \mr{CodeGen comprises four autoregressive language models of varying sizes (350M, 2B, 6B, and 16B). In this paper, we fine-tuned the \textit{Salesforce/codegen-350M-multi} and \textit{Salesforce/codegen-2B-multi} models, each pre-trained on a dataset spanning six programming languages: \texttt{C}, \texttt{C++}, \texttt{Go}, \texttt{Java}, \texttt{JavaScript}, and \texttt{Python}. We selected these models for the TOG task due to their strong code comprehension and generation capabilities.}

\smallskip

\noindent\textbf{PolyCoder~\cite{xu2022systematic}}: \mr{PolyCoder is a family of large language models available in three parameter sizes—400M, 2.7B, and 16B—trained on 249 GB of code across 12 programming languages. In this work, we fine-tuned the \textit{NinedayWang/PolyCoder-0.4B} and \textit{NinedayWang/PolyCoder-2.7B} models for oracle generation.}

\smallskip

\noindent\textbf{Phi-1~\cite{phi1model}}: Phi-1 is a 1.3\mr{-billion-parameter,} decoder-only transformer model \mr{originally designed for} Python code generation. \mr{The \textit{microsoft/phi-1} variant} was pre-trained on Python code from \mr{multiple} sources. Although its primary focus is Python, our study found that it \mr{performs} comparably to other models when fine-tuned for Java code \mr{generation. Nonetheless,} despite achieving good results after fine-tuning, the model struggled to generalize effectively to unseen Java code.

\smallskip

\noindent\textbf{CodeLlama~\cite{codellama7b}}: Code Llama is a family of \mr{pre-trained} and fine-tuned generative models, \mr{ranging in size} from 7 billion to 34 billion parameters. \mr{In our study, we used the \textit{codellama/CodeLlama-7b-hf} model—a 7-billion-parameter, decoder-only language model designed for general code synthesis and comprehension. We selected Code Llama for its strong performance on Java code infilling tasks, where it has outperformed larger models~\cite{roziere2023code}.}

\smallskip

\noindent\textbf{CodeGemma~\cite{codegemma7b}}: \mr{CodeGemma is a collection of lightweight models designed for tasks such as code completion, generation, natural language understanding, mathematical reasoning, and instruction following.} The collection includes a \mr{7-billion-parameter} model pre-trained for code completion and generation, a \mr{7-billion-parameter} instruction-tuned model for code chat and \mr{instruction following}, and a \mr{2-billion-parameter} variant optimized for quick code completion. \mr{In this study, we fine-tuned both the 2-billion and 7-billion pre-trained models (\textit{google/codegemma-2b} and \textit{google/codegemma-7b} from Hugging Face~\cite{huggingface2024}) for test oracle generation.}

\subsubsection{Metric} To compute the accuracy of the generated test oracles, \mr{we use \textit{exact match} with the ground truth oracle as our metric, a common approach in machine learning research}~\cite{hossain2024togll, 10.1145/3660773}. However, we acknowledge that this approach may \mr{\textit{underestimate} accuracy, since a test oracle can be correct even if it does not match the ground truth exactly. For example, we observed assertion oracles like} \texttt{assertFalse(boolean0)} and \texttt{assertEquals(false, boolean0)}, which are semantically equivalent despite \mr{differing} in form.

\subsection{Unseen Inference}
To investigate the generalizability of \mr{\doc comments’ impact on TOG, we perform an unseen inference experiment.}

\subsubsection{Dataset and Metric}\label{dj} \mr{
For this unseen experiment, we use the widely recognized \textbf{Defects4J} dataset~\cite{just2014defects4j}, which contains 835 bugs from 17 real-world Java projects. Each bug (1) is recorded in the project’s issue tracker, (2) involves source code changes, and (3) is reproducible. The dataset provides both buggy and fixed versions of the code. The fixed version applies a minimal patch that resolves the bug, ensuring all project tests pass, while the buggy version fails at least one test. Additionally, Defects4J includes utilities for generating and evaluating test suites, allowing researchers to confirm whether generated tests pass on fixed versions and detect bugs in buggy versions. This dataset is widely used for evaluating automated test oracle generation and program repair methods~\cite{10.1145/3660773, hossain2023neural, hossain2024togll, endres2024can}.

In our study, we \textit{execute the test cases along with our method-generated test oracles} on both the buggy and the fixed versions of each program. To evaluate bug detection effectiveness, we measure (1) the total number of bugs detected by our method, (2) the unique number of bugs detected by each LLM and prompt pair, (3) both individually and in combination, (4) and, finally, we also compare these results with two baseline approaches: \textit{TOGA} \cite{dinella2022toga} and \textit{nl2postcondition}~\cite{endres2024can}.}

\section{Experimental Study}
We examine several key factors: the effects of \doc components on TOG as stand-alone or complementary input, the contribution of various \doc components to TOG, the \mr{effectiveness} of GPT-generated \doc comments, the role of \doc comments in detecting real-world bugs, and when \doc comments can positively/negatively impact TOG. \mr{We answered following \textit{five} research questions}:

\smallskip

\textbf{RQ1:} \mr{What is the impact of developer-written \doc comments on the performance of different LLMs in TOG?}\mr{\textbf{\textit{In almost all cases, including \doc comments improves TOG performance, often by more than 20\% when the base prompt contains minimal information.}}}

\smallskip

\textbf{RQ2:} What is the \mr{contribution of each \doc comment component to the accuracy of TOG?}\mr{\textbf{\textit{We observe that the description and \texttt{@return} tags provide the most value when using \doc comments for TOG.}}}

\smallskip

\textbf{RQ3:} What is the impact of GPT-generated \doc comments on \mr{TOG performance?}\mr{\textbf{\textit{Even GPT-generated \doc comments can improve performance by about 10\% compared to using no \doc comments.}}}

\smallskip

\textbf{RQ4:} What is the impact of \doc comments in detecting \mr{real-world bugs} from Defects4J?
\mr{\textbf{\textit{When \doc comments are included, our method significantly outperforms existing SOTA approaches. Notably, using \doc comments alone can yield better results than using the MUT implementation.}}}

\smallskip

\mrnew{\textbf{RQ5:} What aspects of Javadoc comments make them effective or ineffective for TOG? \textit{\textbf{Javadoc comments are effective when they clearly specify expected behavior, input/output contracts, and return values, but ineffective when they are lengthy, vague/incomplete, or irrelevant}}.}

%revised upto this

\subsection{RQ1: Impact of \doc Comments on TOG}

\mr{This study investigates the impact of \doc comments as a standalone or complementary input source for TOG. We evaluate 60 model-prompt combinations to systematically assess \textit{how \doc comments influence TOG performance across different LLMs and prompt contexts}}.

%In this study, we fine-tune various large language models using the dataset where each input sample includes diverse information, such as the code for the method under test (MUT), a test prefix that exercises the MUT, and the associated \doc comments. We design the prompts to specifically examine the impact of \doc comments on \mr{TOG task}. In total, we evaluate 60 different combinations of model and prompt pairs to thoroughly assess \mr{\textit{how \doc comments influence TOG performance for that LLM and contextual information in the prompt}.}

\subsubsection{Experimental Setup} 
\mr{We fine-tune 10 LLMs using the SF110* dataset (detailed in Section~\ref{training-data}), where each input sample includes the method under test (MUT), a test prefix, and associated \doc comments. To isolate the impact of \doc comments, we design three pairs of prompts. In each pair, the only difference between the two prompts is the inclusion of \doc comments (Section~\ref{prompts}).

Fine-tuning is conducted on a GPU setup with four A100 GPUs (40GB each), utilizing PyTorch’s Accelerate for multi-GPU training. To efficiently train models exceeding 1 billion parameters, we use DeepSpeed~\cite{rasley2020deepspeed}. For 7B models (e.g., CodeLlama-7B, CodeGemma-7B), we apply Low-Rank Adaptation (LoRA)}.

%\mr{We conduct this study using 10 LLMs of varying sizes and families (detailed in Section~\ref{fine-tuning}). Our dataset, SF110*, is described in Section~\ref{training-data}. To examine the impact of \doc comments on \mr{TOG}, we designed three pairs of prompts, where each pair differs only in the inclusion of \doc comments. The prompt details are provided in Section~\ref{prompts}}.

%\mr{For fine-tuning, we used a GPU setup with four A100 GPUs, each equipped with 40GB of memory. To leverage multiple GPUs simultaneously, we used PyTorch’s Accelerate library. To expedite the fine-tuning process for models exceeding 1 billion parameters, we utilized DeepSpeed~\cite{rasley2020deepspeed}. For models with 7 billion parameters (e.g., CodeLlama-7B, CodeGemma-7B), we applied Low-Rank Adaptation (LoRA) to fine-tune only a small subset of parameters. LoRA was configured with a rank $r=16$,  an alpha value of 64, and a dropout rate of 0.05 to prevent overfitting during training. Additionally, the bias was set to `none’ to focus adaptation on the model's core weights.}

\begin{table*}[h]

\footnotesize\centering
\caption{Percentage of ground-truth test oracles generated by various model-prompt combinations. Within each prompt pair, we highlight the best-performing prompt in bold.}
\label{tab:rq1}

%\begin{tabular}[t]{l|c|c|c|c|c|c}
%\toprule
\resizebox{.82\textwidth}{!}{
\begin{tabular}[t]{l|>{\columncolor{col2}}c|>{\columncolor{col2}}c|>{\columncolor{col4}}c|>{\columncolor{col4}}c|>{\columncolor{col6}}c|>{\columncolor{col6}}c}
\toprule
{\diagbox[width=10em]{\thead{\textbf{LLM}}}{\textbf{\thead{Prompt \\Includes:}}}} & \textbf{\thead{Test Prefix}} & \textbf{\thead{Test Prefix,\\ Doc}}  & \textbf{\thead{Test Prefix, \\MUT Sig}} & \textbf{\thead{Test Prefix, \\MUT Sig, \\Doc}} & \textbf{\thead{Test Prefix, \\MUT}} & \textbf{\thead{Test Prefix, \\MUT, \\Doc}} \\
\midrule
 
CodeGPT-110M & 56.80 & \textbf{76.16} & 77.28 & \textbf{78.28} & 79.83 & \textbf{80.07}\\
%1578 & 2115 &  2147 & 2174 & 2162 & 2150
%56.4 & 75.75 & 76.63 & 77.93 & 80.27 & 80.28  \\

CodeParrot-110M & 58.96 & \textbf{78.93} & 79.58 & \textbf{80.23} & 80.48 &  \textbf{80.88}\\
%1638 & 2192 & 2211 & 2228 & 2215 & 2221
%57.84 & 77.6 & 78.43 & 79.04 & 80.82 & 81.27 \\

CodeGen-350M & 58.06 & \textbf{78.57} & 79.62 & \textbf{80.26} &  80.42 &  \textbf{81.56}\\
%1613 & 2182 &   2212 & 2229 &  2211 & 2226
%57.59 & 77.78 & 78.86 & 79.32 & 80.5 & 81.8\\

PolyCoder-400M & 57.63 & \textbf{77.49} &  78.11 & \textbf{78.46} & 79.15 & \textbf{80.67}\\
%1601 & 2152 & 2170 & 2179 & 2146 & 2163
%57.58 & 77.37 & 78.05 & 78.38 & 79.28  & 80\\

Phi-1 & 58.42 & \textbf{78.46} & \textbf{80.16} & 79.15 &  79.73 &  \textbf{80.61}\\
%1623  & 2179 &  2227 &  2198  &  2192  &  2200
%57.77 & 77.74 & 79.44 & 79.57 & 80.28 & 81.16\\

CodeGen-2B & 57.34 & \textbf{78.57} & \textbf{79.12} & 79.04 & 79.59 & \textbf{80.62}\\
%1593 & 2182 & 2198 & 2195 & 2188 & 2200
%57.30 & 78.75 & 79.04 &  79.39 & 80.20 & 81.49 \\

CodeGemma-2B & 58.35 & \textbf{76.84} & \textbf{78.72} & 78.42 & 79.11 & \textbf{80.20}\\
%1621 & 2134 & 2187 & 2178 & 2175 & 2196
%58.13 & 76.88 & 78.68 & 78.17& 79.33 & 80.66\\

PolyCoder-2.7B & 56.04 & \textbf{77.60} & \textbf{78.65} & 78.28 & 78.90 & \textbf{80.23}\\
%1557 & 2155 & 2185 & 2174 & 2139 & 2151
%56.04 & 77.64 & 78.65 &  78.39 & 79.77 & 80.60 \\

CodeLlama-7B & 58.38 & \textbf{79.22} & 78.90 & \textbf{79.07} & 80.75 & \textbf{81.61}\\

%1622 & 2200 & 2192 & 2196 & 2216 &  2219
%58.35 & 79.37 & 78.86 &  79.04& 81.19 & 81.56 \\
CodeGemma-7B & 59.53 & \textbf{79.25} &  79.66 & \textbf{80.41} & 81.08 & \textbf{81.84}\\
%1654 & 2201  & 2213 & 2233 & 2229 &  2241
%59.50 & 79.15  & 79.55 &  80.27 & 80.95 & 81.63\\
\midrule
Average: & 57.95 & 78.11 & 78.98 & 79.16 & 79.90 &  80.83\\
Std. Dev.:	& 0.98	& 0.99	& 0.80	& 0.81	& 0.71	& 0.60\\
\midrule  
t-test (p-value): & \multicolumn{2}{c|}{\textbf{$1.1*10^{-19}$(s)}} & \multicolumn{2}{c|}{0.64 \textbf{(n.s.)}} & \multicolumn{2}{c}{0.00798\textbf{(s)}} \\  
\bottomrule
\end{tabular}}
\end{table*}

\subsubsection{Results}

\mr{The results are presented in Table \ref{tab:rq1}. Column 1 lists the names of all 10 models, arranged in ascending order by parameter size.} Column 2 \mr{shows the TOG} accuracy when only the Test Prefix is used in the prompt ($\mathcal{P}_{1}$), providing a baseline for model performance with minimal contextual information. Column 3 \mr{shows} the accuracy when \doc comments is included in the prompt ($\mathcal{P}_{2}$). Columns 4 and 5 \mr{show} the accuracy for prompts $\mathcal{P}_{3}$ and $\mathcal{P}_{4}$, \mr{both of which incorporate the test prefix and the signature of the MUT}. \mr{ However}, $\mathcal{P}_{4}$ also includes \doc comments. Finally, prompts $\mathcal{P}_{5}$ and $\mathcal{P}_{6}$ utilize the \mr{implementation} of the MUT, \mr{with} $\mathcal{P}_{6}$ additionally \mr{including} \doc comments. Rows 12 and 13 show the average accuracy and the standard deviation for each prompt \mr{across all models, respectively.} \mr{The last row shows the t-test results for each prompt pair, indicating whether adding \doc comments significantly improved the results}.

The average accuracy for $\mathcal{P}_{1}$ is 57.95\%, and \mr{\textit{including \doc comments improves the accuracy by 20 percentage points (pp), increasing it to an average of 78.11\%.} When examining how TOG accuracy shifts with the inclusion of \doc comments ($\mathcal{P}{1}$->$\mathcal{P}{2}$), we found that, on average, across CodeParrot-110M, CodeGen-350M, and CodeGemma-7B, 23\% of samples shift to match, while only 3.3\% shift to no match.}

%The Sankey diagrams in Figure \ref{fig:RQ1} illustrate the shifts in TOG accuracy as \doc comments are included with $\mathcal{P}_{1}$. Subfigures \ref{fig:sub1},  \ref{fig:sub2} and \ref{fig:sub3} show accuracy shifts for CodeParrot-110M, CodeGen-350M and CodeGemma-7B, respectively. For CodeParrot-110M, \doc comments cause 619 samples (22.3\%) to shift from "no match" to "match" with the ground truth oracles, while only 2.7\% move in the opposite direction. This trend is consistent across all models, with an average 23\% shift to match and only 3.3\% shifting to no match.}

%================start of MR part==============
\mrnew{Between $\mathcal{P}_{2}$ and $\mathcal{P}_{3}$, we observe that the method signature (MUT Sig) can be as effective as \doc comments in TOG. To further investigate, we conduct both qualitative and quantitative analyses of the oracles generated using both prompts.

Theoretically, \textbf{\doc comments and method signatures serve distinct purposes, though they may contain overlapping information, such as type details and hints about method functionality derived from the MUT name (e.g., getColumnCount()). A method signature provides only structural details—such as input parameters and return types—but lacks the semantic meaning necessary for understanding intent}. For example, a signature may indicate that a method returns a \textbf{boolean} value, but it \textit{does not} specify under what conditions it returns \texttt{true} or \texttt{false}. This context—essential for accurate reasoning about a method's behavior— can \textit{only} be conveyed through a well-written \doc comment. Figure \ref{fig:rq1-MR} presents three examples demonstrating that method signatures \textit{alone} are \textit{insufficient}. In Example (\textbf{ID: 47331}), the MUT signature indicates a \texttt{boolean} return type, but this information is \textit{insufficient} to determine the expected outcome for a given input.
In contrast, the corresponding \doc comment provides crucial behavioral details: \textbf{the method displays all directories and files with the \texttt{.db} suffix and returns \texttt{true} if the file should be displayed}. The input from the test prefix does not have a \texttt{.db} suffix, so it should not be displayed, and thus, the return value should be \texttt{false}. Utilizing this \doc comments, the model generated a correct oracle, whereas the MUT signature alone failed to infer the expected outcome. The same pattern holds for the other two examples.}

\begin{figure}
  \begin{adjustbox}{width=\textwidth}
\begin{tabular}{>{\centering\arraybackslash}m{4.4cm} >{\centering\arraybackslash}m{4.4cm} >{\centering\arraybackslash}m{4.5cm}}
\begin{lstlisting} [belowskip=-0.9 \baselineskip,style=listingstyle,basicstyle=\ttfamily\tiny,breaklines=true,frame=none]

ID: 47331
===================
MUT Sig: public boolean accept(File f)
----------------------
* The method allows the display of all directories and all files with the *.db suffix.
* @return boolean true if the file should be displayed.
----------------------
//w/ doc.: assertEquals(false, boolean0) // because the file name has no *.db suffix.
////w/ Sig.: assertEquals(true, boolean0);
\end{lstlisting} &  \begin{lstlisting} [style=listingstyle,basicstyle=\ttfamily\tiny,breaklines=true,frame=none,]

ID: 8267
===================
MUT Sig: public static String valueOf(Object value)
----------------------
* Return a empty String if value is null, else return value
* @param value The string value
* @return The result String
----------------------
//w/ doc.: assertEquals("""", string0);
////w/ Sig.: assertNull(string0);
     
\end{lstlisting} & \begin{lstlisting} [style=listingstyle,basicstyle=\ttfamily\tiny,breaklines=true,frame=none]
ID: 64241
===================
MUT Sig: public boolean isDoorOrWindow()
----------------------
* Returns always <code>true</code>.
----------------------
//w/ doc.: assertEquals(true, boolean0)
////w/ Sig.: assertEquals(false, boolean0);
\end{lstlisting}
\end{tabular}
\end{adjustbox}
\vspace{-4mm}
\caption{\mrnew{Examples showing that MUT Sig is not enough to generate correct oracles.}}
\label{fig:rq1-MR}
\end{figure}

\mrnew{We manually analyzed all cases where \doc comments enabled the generation of ground truth test oracles, whereas the MUT signature failed to do so. Our analysis reveals that \doc comments play a crucial role in defining method behavior: in 51.8\% of cases, they explicitly specify the method’s input and expected output, establishing a clear contract \textbf{(input/output)}. In 22.8\%, they emphasize what the method returns, helping in understanding the \textit{expected} output \textbf{(return)}. Additionally, 15.6\% describe the method’s overall functionality without detailing specific inputs or outputs \textbf{(functionality)}, while 9.6\% provide general descriptions that lack focus on return or input/output behavior \textbf{(description)}. These findings highlight that \textbf{well-written \doc comments provide crucial information for TOG that MUT signature alone fail to convey}. 

While our RQ1 study filtered out samples with \textbf{empty \doc comments}, random sampling revealed significant variation in their length and quality across the dataset. Notably, 9\% of \doc comments associated with mismatched oracles were shorter than 42 characters, including vague/irrelevant documentation such as `\texttt{/* Bean Method */}', `\texttt{/ @return */}', `\texttt{/* Default ctor */}', `\texttt{/* {@InheritDoc} */}', and `\texttt{/* Document me */}'. Because the models fine-tuned on $\mathcal{P}_{2}$ were trained to focus on \doc comments, poor-quality comments negatively impacted accuracy (more discussion in RQ5). 
However, this only indicates that \textbf{developer-written \doc comments are \textit{not always well-constructed}—\textit{not} that \doc comments themselves are ineffective}}.

\mrnew{We further analyzed 20 randomly selected cases where \doc comments \textit{did not} match the ground truth, but the method signature did. After carefully reviewing the test prefix, MUT, and \doc comment, we categorized the results into five cases: \textbf{(Case 1)} 7 oracles were \textit{incorrect}, \textbf{(Case 2)} 3 were \textit{correct and equivalent} to the ground truth, \textbf{(Case 3)} 2 were \textit{correct but weaker} than the ground truth,  \textbf{(Case 4)} 1 was \textit{correct and stronger} than the ground truth, and  \textbf{(Case 5)} 7 were \textit{correct but incomparable} to the ground truth.  \textbf{An example of case 4 (ID: 22119)} involves a \doc comment stating, ``This accessor method returns a reference to the live list, not a snapshot". The ground truth oracle is `\texttt{assertNotNull(list1)}', whereas the oracle generated from the \doc comments was `\texttt{assertSame(list1, list0)}', where both lists are derived from calls to the MUT. Here, the LLM leveraged the semantics of \textbf{"live list"} and \textbf{"not a snapshot"} to infer that the MUT is \textbf{idempotent}, leading to a \textbf{stronger} test oracle. \textbf{Overall, among these 20 oracles generated using \doc comments that did not match the ground truth, 13 were correct and capable of effective bug detection. In RQ4, we conducted a bug detection study using real-world bugs, and our results show that oracles generated with only \doc comments can be \textit{even more effective} than those based on the entire MUT implementation in detecting bugs.}}

%Based on this investigation, we calculated the length of the \doc comments and whether they defined the return value (by searching for the word "returns" or the "@return" tag) for cases where either "Doc" or "Sig" matched but not both.  When "Doc" generates a match but "Sig" does not, the \doc comments have a length of (mean = 186.53, median = 131.00, min = 31) and include a return description 59\% of the time.  When "Doc" does not match but "Sig" does, the \doc comments have a length of (mean = 149.89, median = 107.00, min = 18) and include a return description 48\% of the time.  \textbf{This analysis shows that when "Doc" performs better, the documentation is longer (mean +24.4\%, median +22.4\%, min +72\%) and describes the return value 22.9\% more frequently}.
%================end of MR part==============

\mr{Between prompts $\mathcal{P}_{3}$ and $\mathcal{P}_{4}$, including \doc comments also improves accuracy, with a similar trend observed between $\mathcal{P}_{5}$ \textbf{and} $\mathcal{P}_{6}$ (p-values for the t-test are shown in the last row of Table \ref{tab:rq1}). \textbf{Notably, minimal information—specifically, prompt} $\mathcal{P}_{2}$\textbf{—yields performance close to that of the maximum-information prompt} $\mathcal{P}_{6}$. While leveraging MUT code can benefit TOG, it also has inherent limitations: models can inadvertently \textit{learn buggy behaviors from buggy code}, resulting in false negative test oracles that fail to detect bugs, as illustrated in Figure~\ref{fig:example}. \textbf{Findings from this study demonstrate that using only \doc comments—without MUT—can still achieve nearly the same performance}. The MUT and \doc serve as \textit{independent sources of program behavior}, and using them in separate prompts enables independent oracle generation, which can \textit{help identify inconsistencies or conflicts between the MUT and \doc}, as shown in Figure \ref{fig:example}}.  

%\mrnew{In summary, RQ1 reveals through our large-scale qualitative and quantitative study that \doc comments can be extremely useful when well-constructed. However, developer-written comments are often vague and uninformative, making them less effective in such cases. Even when oracles do not match the ground truth, a significant portion are still correct and effective for bug detection. Combining these findings with RQ4, we demonstrate that \doc comments alone can encode sufficient contextual information to replace the MUT, achieving comparable or even superior performance. In Section 4, we discuss best practices for writing high-quality \doc comments to maximize their effectiveness in automated TOG.}

% Key findings
% - 

\begin{tcolorbox}
\textbf{RQ1 Finding:} Including \doc comments across all three prompt pairs provides additional context about the MUT’s behavior, improving TOG to better align with ground truth oracles. Notably, using \doc alone ($\mathcal{P}{2}$) achieves performance comparable to the maximum-information prompt ($\mathcal{P}{6}$). In 74\% of cases where \doc outperformed the MUT signature, \doc comments contained substantive input/output and return value details, highlighting the impact of documentation quality on TOG effectiveness.
For non-matching \doc-generated oracles, 65\% were still correct and capable of detecting bugs, suggesting that an `exact match' with ground truth oracle may underestimate the value of \doc comments.
\end{tcolorbox}

\subsection{RQ2: Impact of Different \doc Components on TOG}
%In RQ1, we conducted a large-scale study using 10 different LLMs pre-trained on code, evaluating their performance across three prompt pairs to investigate the effect of \doc comments on test oracle generation accuracy. \mr{We measured accuracy based on the exact match correctness against the ground truth oracles in the dataset. Our results showed that in nearly 86\% of cases, including \doc comments enhanced the accuracy of generated oracles. For example, adding \doc comments to the test prefix ($\mathcal{P}_{2}$) increased accuracy by almost 20\%.}

\mrnew{A \doc comment can include various components, such as a general description of the MUT, parameter, return, throws, and see tags, along with metadata like version and author details. These comments can be lengthy, and since LLMs have token limits, a key question arises: which \textbf{components} contribute most to improving test oracle generation?

Beyond token limits, writing detailed \doc comments can be time-consuming for developers. Identifying the most impactful components can help streamline documentation efforts while maximizing automated TOG effectiveness.} 

\mr{To address this, in RQ2, we systematically remove different \doc components to assess their impact on TOG accuracy.}

\begin{table*}[t]

\footnotesize\centering
\caption{Percentage of ground-truth oracles generated \mr{with different versions} of \doc comments.}\label{tab:rq2}
\resizebox{.82\textwidth}{!}{
\begin{tabular}[t]{l|c|c|c|c|c|c|c}
\toprule
{\diagbox[width=8em]{\thead{\textbf{LLM}}}{\textbf{Accuracy}}} & \textbf{\thead{Original\\ Javadoc}} & \textbf{\thead{Removed \\Desc.}} & \textbf{\thead{Removed \\@param}} & \textbf{\thead{Removed \\@return}}  & \textbf{\thead{Removed \\@throws}} & \textbf{\thead{Removed \\@see}} & \textbf{\thead{Removed \\Desc. + \\@return}}\\
\midrule

CodeGPT-110M & \textbf{78.27} &  68.56 & 77.26 & 72.80 & 78.23 & 77.91& 54.6\\
%CodeGPT-110M & 75.75 & 66.3 & 74.43 & 70.36 & 75.5\\
CodeParrot-110M & 80.39 & 70.53 & 80.0 & 74.78 & \textbf{80.46} &  80.0 & 55.82\\
%CodeParrot-110M & 77.6 & 68.25 & 77.16 & 72.13 & 77.7\\
CodeGen-350M & 78.48 & 69.17 & 77.87 & 73.66 & \textbf{78.59}& 77.94 & 55.21\\
%CodeGen-350M & 77.78 & 68.53 & 77.32 & 73.16 & 78.19 \\
PolyCoder-400M & \textbf{77.37} & 66.76 & 77.12 & 70.75 & 77.23 & 76.72 & 52.66\\
%PolyCoder-400M & 77.17 & 66.29 & 76.61 & 70.60 & 76.56 \\
Phi-1-1.3B & \textbf{78.38} & 67.37 &  78.12 & 72.80 & 78.38&  77.84& 53.63\\
%Phi-1-1.3B & 79.33 & 67.92 & 78.87 & 73.70 & 79.24 \\
CodeGen-2B & \textbf{78.34} & 68.95 & 78.09 & 73.09 & 78.30 &  77.8 & 53.95\\
%CodeGen-2B & 78.75 & 69 & 78.37 &73.16 & 78.33\\

CodeGemma-2B & \textbf{76.65} & 68.30 & 76.43 & 72.19 & 76.40 & 76.33 & 53.48\\
%CodeGemma-2B & 76.88 & 68.32 & 76.58 &  72.34& 76.53 \\

PolyCoder-2.7B & \textbf{77.51} & 67.15 & 76.90 &72.58 & 77.33&  76.9 & 52.98\\
%PolyCoder-2.7B & 77.64 & 67.37 & 77.18 & 72.63 & 77.89\\

CodeLlama-7B & 79.13 & 67.41 & 78.20 & 73.05 & \textbf{79.24} &  78.66 & 54.78\\
%CodeLlama-7B & 79.37 & 67.27 & 78.16 & 73.21 & 79.35 \\

CodeGemma-7B & 79.17 & 67.05 & 78.59 & 73.52 & \textbf{79.20}&  78.63& 55.0\\
%CodeGemma-7B & 79.15 & 67.06 & 78.53 &   73.56& 79.12 \\
\midrule
\textbf{Avg:} & \textbf{78.37} & \textbf{68.12} & \textbf{77.86} & \textbf{72.92} & \textbf{78.34} &  \textbf{77.87} & \textbf{54.12}\\
\bottomrule
\end{tabular}
}
\end{table*}

\subsubsection{Experimental Setup} \mr{We identified the most frequent \doc components—description, @param, @return, @throws, and @see—and created six modified versions of the \doc comments by removing them individually and removing the description and @return together. We then performed inference using prompt $\mathcal{P}_{2}$ with the test prefix and modified \doc comments to assess their impact on TOG.  \mrnew{This
experimental design allowed us to study the impacts of the
qualitative observations from RQ1 in a more controlled fashion.}}

\begin{figure}
  \begin{adjustbox}{width=\textwidth}
\begin{tabular}{>{\centering\arraybackslash}m{4.4cm} >{\centering\arraybackslash}m{4.4cm} >{\centering\arraybackslash}m{4.5cm}}
\begin{lstlisting} [belowskip=-0.9 \baselineskip,style=listingstyle,basicstyle=\ttfamily\tiny,breaklines=true,frame=none]

//Example: 1
===================
"/**
 * Returns a string representation of this type.
 */"
----------------------
public void test7() {
 Type t0 = Type.getObjectType("[B]");
}
----------------------
//w/ desc.: assertEquals("[B]", t0.toString());
////w/o desc.: assertEquals("[B]", t0.getDescriptor())

\end{lstlisting} &  \begin{lstlisting} [style=listingstyle,basicstyle=\ttfamily\tiny,breaklines=true,frame=none,]

//Example: 2
===================
"/**
* Method getColumnCount
*/"
----------------------
public void test3() {
  PrimaryKey pKey = new PrimaryKey();
  Column c0 = new Column();
  pKey.addColumn(c0);
  boolean bool0 = pKey0.isValid();
}
----------------------
//w/ desc.: assertEquals(1, pKey.getColumnCount());
////w/o desc.: assertEquals(true, bool0);
     
\end{lstlisting} & \begin{lstlisting} [style=listingstyle,basicstyle=\ttfamily\tiny,breaklines=true,frame=none]
//Example: 3
===================
"/**
* Get the time in milliseconds
....
* Default value is 1000 (1 second);
*/"
----------------------
public void test9() {
 FBEventManager fM = new FBEventManager();
 long l0 = fM.getWaitTimeout();
}
----------------------
//w/ desc.: assertEquals(1000L, l0); 
////w/o desc.: assertEquals(0L, l0);
\end{lstlisting}
\end{tabular}
\end{adjustbox}
\vspace{-6mm}
\caption{\mr{Impact of \doc comment's \textit{description} on test oracle generation}}
%Removing the \doc comment description impacting the generated assertion oracles.}
\label{fig:rq2-1}
\end{figure}

\subsubsection{Results}
\mr{The results are presented in Table \ref{tab:rq2}. With the original \doc comments, the average TOG accuracy is 78.37\%. Removing the description has the most significant impact, reducing accuracy to 68.12\%—a 10 percentage points (pp) drop. We manually investigated the 295 affected samples and found that \textit{the description often contains unique information about the MUT that other \doc comment tags do not cover, directly impacting the accuracy of generated oracles}}.

\mr{Figure \ref{fig:rq2-1} presents three examples, each showing the original \doc comments (\textcolor{codeblue}{light blue}), the test prefix, and assertions oracles generated \textit{with} (\textcolor{green}{green}) and \textit{without} (\textcolor{red}{red}) the description. In each case, the description clearly outlines the MUT’s behavior and provides hints for \textit{what to check} and \textit{what to check against}. For instance, in Example 3, the description 
specifies that the oracle should check the  \textit{wait timeout} and default expected value should be 1000 milliseconds.}

\begin{figure}
  \begin{adjustbox}{width=\textwidth}
\begin{tabular}{>{\centering\arraybackslash}m{6.15cm} >{\centering\arraybackslash}m{6.2cm} }

\begin{lstlisting} [belowskip=-0.9 \baselineskip,style=listingstyle,basicstyle=\ttfamily\tiny,breaklines=true,frame=none]

//Example: 1
===================
"/**
* Get the L Norm used.
* @return the L-norm used
*/"
----------------------
public void test1() {
   NaiveBayesMultinomialText nBText = new NaiveBayesMultinomialText();
   String s0 = nBText.stopwordsTipText();
} 
----------------------
//w/ @return: assertEquals(2.0, nBText.getLNorm(), 0.01D);
////w/o @return: assertEquals(2.0, nBText.getLNorm(), 0.01D);
     
\end{lstlisting} & \begin{lstlisting} [style=listingstyle,basicstyle=\ttfamily\tiny,breaklines=true,frame=none]

//Example: 2
===================
"/**
* @return Returns the selectFetchSize.
*/"
----------------------
public void test14() {
    DBCopyPreferenceBean pBean = new DBCopyPreferenceBean();
    pBean.setUseFileCaching(false);
} 
----------------------
//w/ @return: assertEquals(1000, pBean.getSelectFetchSize());
////w/o @return: assertNull(pBean.getName()); 

\end{lstlisting}
\end{tabular}
\end{adjustbox}
\vspace{-4mm}
\caption{\mr{Examples where removing \textit{@return} tag affects (right) and does not affect (left) the generated oracle.}}
\label{fig:rq2-2}
\end{figure}

When the \mr{@return} tags are removed, the average accuracy \mr{decreases} by 5.45 pp. We \mr{\textit{observed a significant overlap between the description and the @return tag in the \doc comments, which helps explain why removing all @return tags results in only a 5.45 pp drop in accuracy}.} Figure \ref{fig:rq2-2} presents two examples: one where removing the @return tag \mr{\textit{affects}} the generated oracle and one where it \mr{\textit{does not}.} For example 1 (left), even though we remove "@return the L-norm used" \mr{portion} from the \doc comments, the description still conveys this information. As a result, removing the @return tag does not \mr{impact} the generated assertion. \mr{Conversely, in example 2 (right), removing the tag directly affects the oracle, as no other part of the \doc comments provides this information.} \mrnew{@return tags provide key hints, specifying \textit{what to check, what to check against, and their types}. In Figure \ref{fig:rq2-2} (Examples 1 and 2), they guide the verification of \texttt{L-norm} and \texttt{selectFetchSize}, directly helping in assertion oracles generation}. 

%\textbf{This analysis offers valuable insights for improving documentation to enhance TOG.}}

\mr{To further investigate this overlap, we removed both the description and the @return tags, affecting 10\% of the samples (288). Removing the description alone caused a 10 pp accuracy drop, while removing the @return tags alone resulted in a 5 pp drop. Combined, this led to a 25 pp accuracy drop, as shown in Column 8 of Table \ref{tab:rq2}. \textbf{In nearly all 288 affected cases, the description and return tag contained the same information, providing the same hints for oracle generation.}}

%To \mr{further} investigate the overlap between the description and the @return tags, we removed both from the \doc comments. We observed that in approximately 10\% of the samples (288), removing both the description and the @return tag \mr{impacted} the generated oracles. Previously, we found that removing the description \textit{\mr{alone}} caused a 10\% accuracy drop, while removing the @return tag \mr{\textit{alone}} resulted in a 5\% drop. Therefore, the combined removal \mr{leads to} a total accuracy drop of 25\%, \mr{as reflected} in the results shown in Column 8. Upon \mr{manually investigating} the 10\% of cases where removing \mr{\textit{both} elements} impacted the generated oracles, we found that in nearly all instances, the description and the @return tag contained the same information. \mr{Often, these tags included key expressions that are used in the generated oracles,} as illustrated in Figure~\ref{fig:rq2-3}. In example 1, both the description and the @return tag document \mr{the return value and type of the MUT.} \mr{Consequently, the generated assertion oracle checks the size of the bin using \texttt{do.getBins()}.} However, when both the description and @return tags are removed, the assertion instead checks the name of the object \texttt{d0}.

\begin{figure*}
  \begin{adjustbox}{width=\textwidth}
\begin{tabular}{>{\centering\arraybackslash}m{6.15cm} >{\centering\arraybackslash}m{6.3cm} }
\begin{lstlisting} [style=listingstyle,basicstyle=\ttfamily\tiny,breaklines=true,frame=none]
//Example: 1
===================
"/**
*  @param  x  The horizontal position of the center of the symbol.
*  @param  y  The vertical position of the center of the symbol.
**/"
----------------------
public void test0() {
    BoxURSymbol symbol0 = new BoxURSymbol();
    symbol0.generatePoints((-2016), (-2016));
    symbol0.generatePoints((-1209), 198);
} 
----------------------
//w/ @param: assertEquals(8, symbol0.getSize());
////w/o @param: exception 
\end{lstlisting} & \begin{lstlisting}[style=listingstyle,basicstyle=\ttfamily\tiny,breaklines=true,frame=none]
//Example: 2
===================
"/**
* @see de.outstare.fortbattleplayer.
*                      model.Sector#getHeight()
*/"
----------------------
public void test20() {
    CharacterClass cC0 = CharacterClass.SOLDIER;
    SimpleSector sS0 = new SimpleSector(1838, true, true, 1838, 1838, true, 1838, cC0);
    boolean boolean0 = sS0.equals((Object) null);
} 
----------------------
//w/ @see: assertEquals(1838, sS0.getHeight());
////w/0 @see: assertEquals(false, boolean0); 
\end{lstlisting}
\end{tabular}
\end{adjustbox}
\vspace{-5mm}
\caption{\mr{Impact of removing @param tag (left) and @see tag (right) on test oracle generation.}}
\label{fig:rq2-4}
\end{figure*}

With the removal of the @param, @throws, and @see tags, we observed that the \mr{\textit{accuracy dropped by less than 1 pp on average}. In examining the 27 samples where removing the @param tag caused the generated oracle to \textit{differ} from the ground truth oracle, we found that the ground truth oracles were more verbose, using fully qualified variable names within the assertion predicates. These fully qualified names were often learned from the @param tag description. \textit{Although these 27 samples did not match the ground truth, the oracles generated without the @param tag were generally more concise}}. In Figure \ref{fig:rq2-4}, we present two examples where the @param and @see tags \mr{provided hints in generating assertion oracles that aligned with the ground truth oracles.}

Regarding the @throws tag removal, \textit{we observed that for 4 out of 10 LLMs, accuracy actually improved slightly, though this improvement was negligible}. \mr{In reviewing some samples}, we found that when the @throws tag \mr{specified} a null pointer exception for null values, the generated assertion oracles often check not null condition using a \texttt{assertNotNull()} oracle. \mr{While these oracles did not exactly match the ground truth, they were still correct.}

\begin{tcolorbox}
\textbf{RQ2 Finding:}
\mr{The \textit{description and @return tags} provide the richest behavioral information about the MUT and are most valuable for automated TOG. These components also significantly overlap. In contrast, @param, @throws, and @see tags contribute little to TOG and can be omitted without significant accuracy loss.}

%Therefore, when \doc comments are lengthy and exceed the maximum token length of LLMs, it is recommended to remove the @param, @throws, and @see tags. Additionally, if a more concise form of \doc comments is desired, and there is significant overlap between the description and @return tags, the @return tag can also be removed without a significant loss in accuracy.
\end{tcolorbox}

%ended here..............
\subsection {RQ3: Assessing GPT-Generated \doc Comments}\label{sec:rq3}

\mr{In RQ1 and RQ2, we investigate the impact of \doc comments and their individual components on TOG, relying on developer-written or derived documentation. However, source code often lacks sufficient documentation—56\% of the SF110 dataset, for example, lacks method-level comments. This raises the question of \textit{whether GPT-generated \doc comments, which summarize MUT behavior, can similarly support TOG}. To explore this, we generate \doc comments directly from the method implementation using a GPT model and evaluate their impact on TOG accuracy.}

\subsubsection {Experimental Setup} For the test portion of the SF110* dataset, consisting of 2,780 input samples, we generate \doc comments using \textbf{OpenAI's gpt-3.5-turbo-0125} model. The generation prompt can be found in our project repository.

%Figure~\ref{fig:rq3-1} presents the generation prompt, which includes a \textbf{system} and a \textbf{user} component. The system component provides model instructions, while the user component supplies the method under test.

%In this study, we \mr{utilize} all fine-tuned models from RQ1 and the test portion of the dataset, which consists of a total of 2,780 input samples, all with \doc comments. \mr{Additionally, we create an alternative version of the test dataset by generating \doc comments for each method using a Generative Pre-trained Transformer (GPT) model, specifically OpenAI's gpt-3.5-turbo-0125 model}. Figure~\ref{fig:rq3-1} shows the generation prompt, which consists of two components: system and user. The system component \mr{provides} the instructions for the model to follow, while the user component \mr{supplies} the method under test.

\begin{table*}[h]
\caption{Percentage of ground-truth test oracles generated with GPT-generated \doc comments.}\label{tab:rq3-2}
\small
\centering
\resizebox{\textwidth}{!}{%
\begin{tabular}[t]{c|c|c|c|c|c|c|c|c|c|c|c}
\toprule
{\diagbox[width=7em]{\textbf{\thead{Prompt}}}{\thead{\textbf{Model}}}} 
& \textbf{\thead{CodeGPT\\110M}} 
& \textbf{\thead{CodeParrot\\110M}} 
& \textbf{\thead{CodeGen\\350M}} 
& \textbf{\thead{PolyCoder\\400M}} 
& \textbf{\thead{Phi\\1.3B}}  
& \textbf{\thead{CodeGen\\2B}} 
& \textbf{\thead{CodeGemma\\2B}} 
& \textbf{\thead{PolyCoder\\2.7B}} 
& \textbf{\thead{CodeLlama\\7B}} 
& \textbf{\thead{CodeGemma\\7B}} 
& \textbf{\thead{Avg.}} \\
\midrule

\thead{ $\mathcal{P}{1}$} 
& 56.80  & 58.96 & 58.06 & 57.63 & 58.42 & 57.34 & 58.35 & 56.04 & 58.38 & 59.53 & \textbf{57.95} \\
\midrule

\thead{$\mathcal{P}{1}$ + \\GPT-generated \\\doc} 
& 66.07 & 70.79 & 69.71 & 64.42 & 67.73 & 70.54 & 67.30 & 67.87 & 66.90 & 66.72 & \textbf{67.80} \\
\bottomrule
\end{tabular}%
}
\end{table*}
\subsubsection {Results} 
\smallskip\smallskip\smallskip

\begin{figure*}[t]
  \begin{adjustbox}{width=\textwidth}
\begin{tabular}{>{\centering\arraybackslash}m{7 cm} >{\centering\arraybackslash}m{7 cm}}

\begin{lstlisting} [belowskip=-0.3 \baselineskip,style=listingstyle,basicstyle=\ttfamily\tiny,breaklines=true,frame=none]
// Example: 1
============================
---GPT-generated Javadoc 
"/**
 * Returns whether the object is coded or not.
 * @return true if the object is coded, false otherwise
 */"
----------------------
public void test14() {
 RawVariable rV = new RawVariable();
 SupportingDocument sDoc = 
     new SupportingDocument();
 boolean b0 = 
     rV.containsSupportingDocument(sDoc);
}
----------------------
// Same as ground truth assertion 
// w/ GPT-doc: assertEquals(false, rV.isCoded());

////w/o GPT-doc: assertEquals(false, rV.isCleaned());
\end{lstlisting} &  \begin{lstlisting} [style=listingstyle,basicstyle=\ttfamily\tiny,breaklines=true,frame=none,]
// Example: 2
============================
---GPT-generated Javadoc 
"/**
 * This method retrieves the current index value.
 * @return The current index value.
 */"
----------------------
public void test0() {
 Object obj = new Object();
 NoteListDataEvent event = new NoteListDataEvent(obj, 2);
 int type = event.getType();
}
----------------------
// Same as ground truth assertion 
// w/ GPT-doc: assertEquals(2, event.getIndex());

////w/o GPT-doc: assertEquals(2, type);
\end{lstlisting} 
\end{tabular}
\end{adjustbox}
\vspace{-4mm}
\caption{Impact of GPT-generated \doc comments on ground truth oracle generation.}
\label{fig:rq3-3}
\end{figure*}

\smallskip\smallskip

In Table~\ref{tab:rq3-2}, we \mr{present} the TOG performance of 10 different fine-tuned models on two \mr{distinct} prompts. The first prompt ($\mathcal{P}{1}$) \mr{includes only} the test prefix, while the second prompt ($\mathcal{P}{1}$ + GPT-generated \doc) \mr{incorporates both} the test prefix and the generated Javadoc comments. \mr{The purpose of using these two prompts is to isolate the impact of GPT-generated \doc comments on TOG. When using $\mathcal{P}{1}$, the average accuracy is 57.95\%. \textit{Incorporating GPT-generated \doc comments increases the average accuracy to 67.8\%, indicating a 10 pp improvement.}}

\mr{From the samples showing improvement (from no match to match) after adding generated \doc comments, we manually investigated 20 cases. When comparing the test oracles generated using $\mathcal{P}{1}$ (test prefix only) to those generated with $\mathcal{P}{1}$ + GPT-generated Javadoc, we observed that the GPT Javadoc provided useful hints that were instrumental in constructing the oracle statements. Figure \ref{fig:rq3-3} illustrates two representative examples of generated \doc comments (\textcolor{codeblue}{light blue}), the assertions generated using only the test prefix (\textcolor{red}{red}), and the assertions generated using both the test prefix and the GPT-generated Javadoc comments (\textcolor{green}{green}). Each \doc comment includes a summary of the MUT and a @return tag, explicitly specifying the method's return behavior.} In RQ2, we established that the method summary and @return tag are the most impactful components for automated test oracle generation. These examples further reinforce this finding.

\begin{tcolorbox}

\textbf{RQ3 Finding:}
We find that \mr{including GPT-generated \doc comments with the test prefix improves TOG} accuracy by 10 pp. This \mr{generalizes} the finding in RQ1 that \doc comment provides \mr{additional} context about MUT behavior, \mr{enhancing the generation of effective test oracles. Moreover, GPT-generated Javadoc comments that contributed to generating ground truth oracles consistently include a method description and a @return tag, further supporting the RQ2 finding that these components are the most valuable for TOG.}
\end{tcolorbox}

%ended here
\subsection{RQ4: Impact of \doc Comments on Real Bug Detection}

In the previous RQs, we examined the correctness of automatically generated test oracles. \mr{While \textit{correctness} is a critical property~\cite{HossainEnsuring}, the \textit{ability to detect bugs—referred to as the strength property}—is equally important}. In this research question, we investigate the impact of \doc comments on bug detection using \textit{Defects4J}, a widely adopted real-world Java bug benchmark. We evaluate multiple fine-tuned LLMs and prompts to compare the relative strengths of the generated oracles and benchmark our results against two state-of-the-art methods.

%In the previous RQs, we investigated the correctness of the automatically generated test oracles. \mr{While the \textit{correctness}} of the test oracles is a \mr{critical} property~\cite{HossainEnsuring}, the \mr{\textit{ability to detect bugs—referred to as the strength property}}—is equally important. In this research question, we \mr{examine} the impact of \doc comments on bug detection using Defects4J, a \mr{widely adopted} real-world bug benchmark for Java programs. We \mr{evaluate multiple fine-tuned} LLMs and prompts to compare the relative strengths of the generated oracles. \mr{Additionally, we compare our results against two baseline state-of-the-art methods.}

\begin{table*}[h]
\caption{Defects4J bug detection performance of model-prompt pairs on the dataset subset (288/374), where each input sample \textit{includes} a \doc comment.}\label{tab:rq4-1}
\resizebox{.7\textwidth}{!}{
\begin{tabular}[t]{c|c|c|c|c|c}
\toprule
{\diagbox[width=8em]{\thead{\textbf{Prompt}}}{\thead{\textbf{Model}}}} & \textbf{\thead{CodeParrot-110M}} & \textbf{\thead{CodeGen-350M}} &\textbf{\thead{Phi-1.3B}}  & \textbf{\thead{CodeLlama-7B}} & \textbf{\thead{CodeGemma-7B}}\\
\midrule

\thead{prefix ($\mathcal{P}{1}$)} & 44 (4) & 44 (4) & 42 (4) & 42 (5) & 45 (4)\\
\midrule
\thead{prefix +\\ \doc comments ($\mathcal{P}{2}$)}& 44 (4) & \textbf{50} (10) & 42 (4) & \textbf{48} (11) &  \textbf{48} (7)\\
\midrule
\rowcolor{col3} \thead{\textbf{Total Bugs ( Unique ) :}}  & 48 &  \textbf{54} & 46 &  \textbf{53} &  \textbf{52} \\

\bottomrule
\end{tabular}
}
\end{table*}
\subsubsection{Experimental Setup} Details of the Defects4J dataset are provided in Section \ref{dj}. Similar to prior works \cite{dinella2022toga, endres2024can, hossain2024togll}, this dataset includes a total of 374 input samples. Each sample comprises a test prefix, a method under test, and its corresponding \doc comments. For the 86 samples where \doc comments were \textit{unavailable}, we generated them using the procedure described in RQ3 (Section~\ref{sec:rq3}).

\mr{Based on RQ1 results, we selected the top five models that performed best on prompt $\mathcal{P}_{2}$ while ensuring diversity by avoiding repetition within the same model family: \textbf{CodeParrot-110M, CodeGen-350M, Phi-1.3B, CodeLlama-7B, and CodeGemma-7B}. 

After an initial study on all five models using $\mathcal{P}{1}$ and $\mathcal{P}{2}$ on the Defects4J dataset (Table~\ref{tab:rq4-1}), we selected the top three for further analysis: \textbf{CodeGen-350M, CodeLlama-7B, and CodeGemma-7B}. Using these models and prompt pairs ($\mathcal{P}{1}$, $\mathcal{P}{2}$) and ($\mathcal{P}{5}$, $\mathcal{P}{6}$), we generated test oracles to detect Defects4J bugs. Table~\ref{tab:rq4-2} reports the total number of unique bugs detected by different model-prompt combinations.}

\subsubsection{Results} \mr{Table~\ref{tab:rq4-1} presents the Defects4J bug detection performance of five models using prompts $\mathcal{P}{1}$ and $\mathcal{P}{2}$. Here, $\mathcal{P}{1}$ includes only the test prefix, while $\mathcal{P}{2}$ includes both the test prefix and \doc comments. This study evaluates 288 input samples, all containing \doc comments. Most models generated stronger test oracles and detected more bugs with $\mathcal{P}{2}$ than with $\mathcal{P}{1}$. This \emph{enhanced bug detection performance can be directly attributed to the \doc comments}, as they are the only distinguishing feature between these two prompts.}

\mr{For example, CodeGen-350M, CodeLlama-7B, and CodeGemma-7B detected more bugs with $\mathcal{P}{2}$, highlighting the value of \doc comments in generating stronger test oracles. CodeGen-350M and CodeLlama-7B detected 10 and 11 unique bugs, respectively, using \doc comments. The last row shows the total number of unique bugs detected when combining $\mathcal{P}{1}$ and $\mathcal{P}{2}$.

For CodeParrot-110M and Phi-1.3B, the inclusion of \doc comments does not improve performance, suggesting that different fine-tuned models generalize differently during unseen inference. CodeParrot-110M, the smallest model in this study, and Phi-1.3B, pre-trained on Python datasets, may have limited generalization capabilities for Java code generation. Despite performing well in RQ1, these models struggle to generalize effectively in the TOG task. Based on these results, we select the top three models—CodeGen-350M, CodeLlama-7B, and CodeGemma-7B—for deeper investigation.}

\begin{table*}[t]
\caption{Defects4J bug detection performance of various model-prompt pairs}\label{tab:rq4-2}
\resizebox{.7\textwidth}{!}{
\begin{tabular}[t]{l|c|c|c}
\toprule
 {\diagbox[width=11em]{\thead{\textbf{Prompt}}}{\textbf{Model}}} & \textbf{\thead{CodeGen-350M}} & \textbf{\thead{CodeLlama-7B}} & \textbf{\thead{CodeGemma-7B}}\\
\midrule

 \thead{prefix ($\mathcal{P}{1}$)} & 59 (5)  & 58 (8)  & 62 (3)   \\
\midrule
 \rowcolor{col3} \thead{prefix + \doc comments ($\mathcal{P}{2}$)}& \textbf{67} (13) & \textbf{63} (13) & \textbf{68} (9) \\
\hline
\thead{\textbf{Total Unique ($\mathcal{P}{1}$+ $\mathcal{P}{2}$):}} & 72  & 71 & 71 \\
\hline
\midrule
\thead{prefix + MUT ($\mathcal{P}{5}$)} & 62 (4) & 68 (5) & 62 (4) \\
\midrule
\rowcolor{col3} \thead{prefix + MUT +\\ \doc comments ($\mathcal{P}{6}$)}& \textbf{63} (5) & \textbf{73} (10) & \textbf{65} (7) \\
\hline
\thead{\textbf{Total Unique ($\mathcal{P}{5}$+ $\mathcal{P}{6}$):}} &  67 & 78 & 69 \\
\bottomrule
\end{tabular}}
\end{table*}

\mr{Table \ref{tab:rq4-2} presents the bug detection results for the top three models across all 374 input samples. This study evaluates two prompt pairs, totaling four distinct prompts. Notably, including all 374 samples in this evaluation resulted in a higher total bug detection count for both prompts compared to Table \ref{tab:rq4-1}.

Row 2 presents bugs detected using $\mathcal{P}_1$ (test prefix), while Row 3 shows results for $\mathcal{P}_2$ (test prefix + \doc comments). With \doc comments, CodeGen-350M, CodeLlama-7B, and CodeGemma-7B detected 8, 5, and 6 additional bugs, respectively. A similar trend is observed between $\mathcal{P}_5$ and $\mathcal{P}_6$. In summary, \doc comments, as a \emph{complementary} source of information, consistently led to the detection of more Defects4J bugs.

Comparing results from $\mathcal{P}_2$ and $\mathcal{P}_5$, we find that \textit{\doc comments alone can encode sufficient contextual information to replace the MUT code entirely, achieving comparable or even superior performance in TOG and bug detection}. $\mathcal{P}_2$ (prefix + \doc comments) detects a maximum of 68 and an average of 66 bugs, while $\mathcal{P}_5$ (prefix + MUT) detects a maximum of 68 and an average of 64 bugs. \textit{This suggests that \doc comments can be used in place of MUT implementation for TOG.}}

\mr{We also find that for 2 of the 3 models, using MUT implementation does not improve bug detection compared to \doc comments ($\mathcal{P}_2$ vs. $\mathcal{P}_6$). A possible reason is that models may struggle to extract high-level behavioral insights from complex MUT code, while \doc comments offer a more accessible, natural-language summary.

In summary, \doc comments, when used as a standalone input, achieved comparable or better performance than MUT code. Thus, they have the potential to replace MUT while maintaining similar performance. The impact of \doc comments \mr{remain}s significant \mr{even} when the code for the MUT is included. When \doc comments \mr{are used} with the prefix and MUT (Row 6), \mr{the total number of detected bugs increases in all cases. Notably, the best} performance is achieved by CodeLlama-7B, which detected a total of 73 bugs. \mr{Furthermore}, the total number of unique bugs \mr{is substantially} higher when \doc comments are included, \mr{highlighting their critical role in TOG and bug detection.}}

\begin{table*}[t]
\Small\centering

\caption{Impact of \doc comments on Defects4J bug detection.}\label{fig:rq4}

\begin{tabular}{>{\centering\arraybackslash}m{2cm} | >{\arraybackslash}m{5cm} | >{\arraybackslash}m{5.6cm}}

\hline
\textbf{Bug ID}  &  \textbf{Cli, 34}  & \textbf{Csv, 6}\\
\hline
\textbf{Bug Details} &
getParsedOptionValue returns null unless Option.type gets explicitly set.

\vspace{1mm}
The user expects it to be String unless set to any other type.

\vspace{1mm}
This could be either fixed in the Option constructor or in CommandLine.getParsedOptionValue. Mentioning this behaviour in Javadoc would be advisable. & 
if .toMap() is called on a record that has fewer fields than we have header columns we will get an ArrayOutOfBoundsException.

\vspace{2mm}
CSVRecord.toMap() fails if row length shorter than header length.\\
\hline
\textbf{Javadoc Comment:} & \begin{lstlisting} [style=listingstyle,basicstyle=\ttfamily\tiny,breaklines=true,frame=none,label={lst:optional}]
/**
* Retrieve the type of this Option. 
*/
\end{lstlisting}&   \begin{lstlisting} [style=listingstyle,basicstyle=\ttfamily\tiny,breaklines=true,frame=none,label={lst:optional},belowskip=-0.8 \baselineskip]
/**
* Returns the number of this record in the 
* parsed CSV file.
*/
\end{lstlisting}\\
\hline
\textbf{Test Prefix:}&\begin{lstlisting} [style=listingstyle,basicstyle=\ttfamily\tiny,breaklines=true,frame=none,label={lst:optional},belowskip=-0.8 \baselineskip]
public void test09() {
  Option op0 = new Option("", "", true, "");
  Object obj0 = op0.getType();
}
\end{lstlisting}&\begin{lstlisting} [style=listingstyle,basicstyle=\ttfamily\tiny,breaklines=true,frame=none,label={lst:optional},belowskip=-0.8 \baselineskip]
public void test17() {
   HashMap<String, Integer> hMap = new HashMap<String, Integer>();
   Integer int0 = new Integer(854);
   hMap.put((String) null, int0);
   String[] strA0 = new String[0];
   CSVRecord cSVRec = new CSVRecord(strA0, hMap, (String) null, 854);
   cSVRec.toMap();
}
\end{lstlisting}\\
\hline
\textbf{ Javadoc 
\textit{not} used:} & \begin{lstlisting} [style=listingstyle,basicstyle=\ttfamily\tiny,breaklines=true,frame=none,label={lst:optional},belowskip=-0.8 \baselineskip]
assertEquals(true, op0.isIsSet()); \end{lstlisting} &  \begin{lstlisting} [style=listingstyle,basicstyle=\ttfamily\tiny,breaklines=true,frame=none,label={lst:optional}]
assertEquals(854, cSVRec.size()) \end{lstlisting}\\
\hline
\textbf{ Javadoc  used:} & \begin{lstlisting} [style=listingstyle,basicstyle=\ttfamily\tiny,breaklines=true,frame=none,label={lst:optional},belowskip=-0.8 \baselineskip] 
assertNotNull(obj0);
\end{lstlisting} & \begin{lstlisting} [style=listingstyle,basicstyle=\ttfamily\tiny,breaklines=true,frame=none,label={lst:optional},belowskip=-0.8 \baselineskip] 
assertEquals(854L, cSVRec.getRecordNumber());
\end{lstlisting}\\
\hline
\end{tabular}
\end{table*}

\mr{Figure \ref{fig:rq4} presents two Defects4J bugs uniquely detected due to \doc comments. For example, bug 34 in the \textit{Cli} project occurs when the \texttt{Option} type is not explicitly set. If undefined, the user expects it to default to a \texttt{String}. Without \doc comments, the generated test oracle passed in both fixed and buggy versions, failing to detect the bug. However, with \doc comments, the assertion ensured the type was not \texttt{null}, passing in the fixed version but failing in the buggy one—successfully detecting the bug. This demonstrates how \doc comments strengthen test oracles by capturing intended behavior not directly inferred from the MUT.

Similarly, in bug 6 of the \textit{Csv} project, \texttt{CSVRecord.toMap()} fails when the row length is shorter than the header length, triggering an \texttt{ArrayOutOfBoundsException}. Without \doc comments, the generated oracle only checks the \texttt{CSVRecord} size, producing an incorrect test that failed to detect the bug. In contrast, with \doc comments, the oracle checks the record number as specified, successfully detecting the bug.}

%\soneya{stopped here..........}

\noindent\textbf{Comparison With Baselines:} We compare our results with two baseline methods. The first is \textit{TOGA} \cite{dinella2022toga}, a neural \mr{approach} for test oracle generation. TOGA \mr{utilizes} the test prefix, MUT, and \doc comments to generate test oracles for detecting Defects4J bugs. \mr{While} TOGA detected a total of 57 bugs, a significant portion of these were \mr{identified using} implicit oracles. \mr{Specifically, when} TOGA failed to generate explicit assertion oracles, it executed the test prefix (not generated by TOGA) on buggy programs, where uncaught exceptions caused test \mr{failures, indirectly indicating bug detection. In contrast, our method does not execute test prefixes when no oracles are generated.} Despite this, \mr{our approach demonstrates} a significant improvement over TOGA, detecting 73 bugs when the same input information is used and 68 bugs even when MUT code is excluded from the input prompt.

We note that using the MUT from the fixed version, as TOGA does, is impractical \mr{because} fixed versions are typically unavailable in real-world scenarios, as demonstrated by \mr{prior} research \cite{liu2023towards, hossain2023neural}. \mr{Moreover}, if the buggy MUT is used, the generated oracles \mr{risk learning} from buggy behavior, \mr{which increases} the likelihood of producing regression oracles that fail to detect bugs in that program version. \mr{Therefore, achieving state-of-the-art bug detection capability using only \doc comments, as demonstrated by our method, offers a viable solution to mitigate these challenges.}

The second \mr{baseline} method is \textit{nl2postcondition}~\cite{endres2024can}, which uses GPT-4 and StarChat to generate test oracles from \doc comments and buggy MUT code. This method also \mr{incorporates} class-level in-file comments, \mr{\textit{which are not utilized in our study}}. For each input sample (\mr{comprising} \doc comments and buggy MUT code), \mr{\textit{nl2postcondition}} generates 10 assertions. If at least one of the generated assertions \mr{passes on} the fixed version and \mr{fails on} the buggy version, \mr{the method reports a detected bug}. In contrast, our study generates only a single test oracle \mr{per} input sample.

\mr{Despite additional context and advantages, \textit{nl2postcondition}—using GPT-4 with \doc comments and buggy MUT code—detected a total of \textbf{47 unique bugs~\cite{endres2024can}}. In contrast, our method, relying solely on \doc comments and generating a single assertion per sample, detected 68 unique bugs—an improvement of nearly 45\%. When the MUT was removed from \textit{nl2postcondition}, its performance dropped significantly, detecting only 35 bugs. This highlights a 94\% improvement by our method, which detected 68 bugs under the same conditions.}

The significant improvement of our approach can be attributed to fine-tuning with a high-quality dataset collected from diverse, real-world projects, well-designed prompts, and careful selection of the most relevant parts of the \doc comments, as determined by our RQ2 findings.

\begin{tcolorbox}
\textbf{RQ4 Finding:} \mr{The most important finding is that \doc comments alone can encode sufficient contextual information to potentially replace the MUT code, achieving comparable or even better performance in TOG and bug detection. Across all models}, prompts with \doc comments detect more bugs, \mr{indicating} that the generated oracles are stronger and better at capturing developer intentions. Our method, without using the MUT, detects 19–45\% more bugs than prior work \mr{that} uses the MUT, and \mr{94\% more bugs when prior methods exclude the MUT.}
\end{tcolorbox}

\mrnew{\subsection{RQ5: Helpful or Harmful Aspects of \doc Comments} }

\mrnew{In this research question, we conduct a comprehensive qualitative and quantitative analysis to identify the aspects of \doc comments that are \textit{helpful} or \textit{harmful} for test oracle generation. First, we present a taxonomy categorizing the reasons why Javadoc can enhance or hinder test oracle generation. Then, we provide concrete examples of cases where adding \doc comments did not improve performance and analyze the underlying reasons.}

\mrnew{\subsubsection{Experimental Setup} To identify \textbf{positive factors}, we extracted cases where \doc comments helped generate a ground truth oracle. For \textbf{negative factors}, we examined instances where \doc comments resulted in oracles that did not match the ground truth. To identify cases where \doc comments had \textbf{no impact}, we analyzed situations where the generated oracle remained \textbf{unchanged}. 
Our prompt design naturally supports these investigations.

We conducted a manual investigation to develop a taxonomy of why \doc comments are helpful or harmful for test oracle generation. This involved systematically analyzing \doc comments, identifying recurring patterns, and categorizing the reasons. We iteratively refined these categories to ensure clear and meaningful distinctions.
}

%We applied \textbf{grounded theory} using an open coding approach to develop a taxonomy of why \doc comments are helpful or harmful. This involved manually defining \textbf{codes} (reasoning categories) based on keywords in the \doc comments and iterating until clear definitions emerged.}  

\mrnew{\subsubsection{Results} To find the positive reasons we have manually investigated a total of \textbf{900} samples across all prompt pairs \textbf{($\mathcal{P}_1$ -> $\mathcal{P}_2$, $\mathcal{P}_3$ -> $\mathcal{P}_4$, $\mathcal{P}_5$ -> $\mathcal{P}_6$)}. We identified \textbf{four} positive reasons why adding \doc comments helped generate a ground truth test oracle:}

\begin{figure}
  \begin{adjustbox}{width=\textwidth}
\begin{tabular}{>{\centering\arraybackslash}m{4.8cm} >{\centering\arraybackslash}m{4.8cm} >{\centering\arraybackslash}m{4.5cm}}
\begin{lstlisting} [belowskip=-0.9 \baselineskip,style=listingstyle,basicstyle=\ttfamily\tiny,breaklines=true,frame=none]

ID: 54958 (Return behavior)
===================
"/**
* Returns the single line comment start string.
* @return the start string
*/"
----------------------
//w/ doc.: assertEquals(""//"", doc.getSingleLineCommentStart());
////w/o doc: assertEquals(2, doc.getIndentationSize());
\end{lstlisting} &  \begin{lstlisting} [style=listingstyle,basicstyle=\ttfamily\tiny,breaklines=true,frame=none,]

ID: 33999 (Input/Output)
===================
"/**
* Removes all null values from the given string array.
* Returns a new string array that contains all none null values of the input array.
* @param strings The array to be cleared of null values
*/"
----------------------
//w/ doc: assertNull(stringArray0);
////w/o doc.: assertNotNull(stringArray0);
\end{lstlisting} & \begin{lstlisting} [style=listingstyle,basicstyle=\ttfamily\tiny,breaklines=true,frame=none]
ID: 141507 (Functionality)
===================
"/**
* Returns the algorithm for the QuickServer used for key management 
* when run in a secure mode.
* @see #setAlgorithm
*/"
----------------------
//w/ doc.: assertEquals(""{_dWSu0rBmFf"", secureStore0.getAlgorithm());
////w/o doc.: something else
\end{lstlisting}
\end{tabular}
\end{adjustbox}
\vspace{-4mm}
\caption{\mrnew{Examples showing positive reasons of \doc comments for effective TOG}}
\label{fig:rq5-MR-1}
\end{figure}

\mrnew{
\begin{description}
    
    \item[Return behavior]{When the  \doc comments explicitly defines the \textbf{return value}, either in the \textbf{description (using "Returns...")} or via the \texttt{@return} tag.}

    \item[Functionality]{When the \doc comments clearly states the \textbf{purpose} or \textbf{expected behavior} of the method.}

    \item[Input/output]{When both method \textbf{parameters} and \textbf{return} values are \textit{explicitly} stated, often including the \texttt{@throws} tag.}
    
    \item[Description]{When the \doc comments provides some descriptive information, though with \textbf{limited} detail.}
    
\end{description}}

\mrnew{
Figure \ref{fig:rq5-MR-1} presents three examples. In \textbf{ID: 54958}, we show how the return behavior described in the \doc comments directly influenced ground truth oracle generation. In this category, \doc comments also include additional details such as \textit{method functionality or general descriptions, but the return behavior played the most significant role in guiding oracle generation}. As a result, we classified these cases under \textbf{Return behavior}. \textit{Return behavior documentation provides critical hints for both "what to check" and "what to compare with"}. For instance, in \textbf{ID: 54958}, the \doc comment specifies that "what to compare with" is the "single line comment". Similarly, the examples in Figure \ref{fig:rq2-2} illustrate "what to check". Notably, this category accounts for \textbf{65.64\% of all positive contributions}, emphasizing the crucial role of well-defined return behavior for TOG.

In example \textbf{ID: 141507}, the \doc comment states that the method returns the algorithm for the QuickServer used for key management when running in secure mode and references \texttt{@see \#setAlgorithm}. This information helped construct the oracle by guiding the model to learn "what to compare with" from the test prefix and "what to check" from the \doc comment. This  \textbf{Functionality} category accounts for 26.93\% of all positive contributions.

In example \textbf{ID: 33999}, the \doc comment describes the input and output behavior of the MUT, aiding in the generation of the ground truth oracle. Since the test prefix provides the input, the model inferred the expected output from the \doc comments. This \textbf{Input/Output} category accounts for 3.05\% of all positive contributions.

The last category, \textbf{Description} only provides very limited context on the MUT, e.g., `\texttt{This method computes the dot product}', or `\texttt{Helper method for encoding an array of bytes as a Base64 String}' with no other information. In the 4.36\% of the cases in our study, the specifics of these descriptions
did help the TOG task -- \textit{perhaps because the LLMs have learned a good embedding of the mentioned
concepts, e.g., `dot product'}.

\textbf{Solving the test oracle problem fundamentally requires understanding the expected behavior of a method for a given input. Therefore, as long as \doc comments convey this information, they contribute to effective oracle generation. While we have categorized the reasons, they ultimately serve the same core purpose—providing guidance on \textit{what to check} and \textit{what to expect}}.

To identify the \textbf{negative} characteristics of \doc comments, we analyzed all 229 samples in which the first prompts in a pair, e.g., $\mathcal{P}_3$, matched the ground truth but the second, e.g., $\mathcal{P}_4$, did not. Among these, 37.1\% were \textbf{correct} oracles that did not match the ground truth. Of the remaining 144 samples, we identified \textbf{three} distinct reasons why \doc comments are \textbf{not helpful}: }

\begin{description}
    \item[Lengthy:]{When the \doc comment was excessively long it could exceed the LLM token limit; this occurred in 3.5\% of the incorrect samples. 
        IDs 100041, 2322, 100332, and 780 all have very detailed input/output descriptions, but 
    the length of prefix+signature+doc \textbf{exceeds the token length}, so no oracle is generated.}

    \item[Vague/Incomplete:]{When the \doc comment contained some \textbf{relevant} information but \textbf{did not explicitly describe all input/output behavior or return values}; this occurred in 55.6\% of the incorrect samples. We describe four randomly selected cases:
    ID 44613 has a description of \texttt{/* isophote flow??? */};
    ID 49847 has a description of \texttt{/* Method getXMLName */}, \textbf{which adds nothing beyond part of the information in the signature}; 
    ID 156278 has a description of \texttt{/* toString means print the data string, unless the data has not been read at all. */}, \textbf{which does not define either input/output behavior or the functionality of the MUT}; and 
    ID 135723 has a description of \texttt{/* {@inheritDoc}*/}}.

    \item[Irrelevant:]{When the \doc comment included text \textbf{unrelated to the MUT}; this occurred in 15.3\% of the incorrect samples. We describe four randomly selected cases:
ID 10004 has a description that references various methods and classes but ends with \texttt{@return Document me!};
    IDs 9790 and 11287 have \doc containing a meaningful \texttt{@return} tag, but the description consists only of \texttt{DOCUMENT ME!}; and
    ID 16507 has the following description:
\begin{small}
\begin{verbatim}
/* Taken from Eammon McManus' blog:
 * http://weblogs.java.net/blog/emcmanus/archive/2007/03/getting_rid_of.html.
 * Prevents the need for placing SuppressWarnings throughout the code. */
\end{verbatim}
\end{small} } 
\end{description}
We categorized samples as \textbf{unknown}
when none of the above categories applied; this occurred in 25.7\% of the incorrect samples.

\mrnew{We further analyzed cases where adding \doc comments \textbf{did not change the results}. From $\mathcal{P}_1$ → $\mathcal{P}_2$, we identified 512 instances where the generated oracles \textbf{remained unchanged (non match with ground truth)}. Among them, 196 had identical oracles for both prompts. \textbf{Upon analysis, we found that 88 were correct, 66 were vague, 22 were irrelevant, 17 were unknown, and 3 were lengthy}. These findings align with the previously discussed \textbf{negative reasons}.}

\mrnew{
In summary, \doc comments can contain excessively long text, vague/insufficient details (e.g., \texttt{@inheritDoc}, \texttt{DOCUMENT ME!}), or consist solely of a \texttt{@see} tag. Additionally, many fail to describe the input/output contract or specify return values. Due to these shortcomings, such \doc comments do not contribute to effective oracle generation. \textbf{We conjecture that lengthy, vague/incomplete, or irrelevant documentation misguides the model, preventing it from learning meaningful behavioral constraints}. Our findings emphasize the need for well-structured documentation, guiding developers to improve \doc comments for effective TOG.
}

%\matt{need to discuss how some of these negatives can combine with positives but lead to a bad result, e.g., length, and ID 11287 are examples.}

\section{Threats to Validity}
In our large-scale study, we leveraged the widely recognized and open-source SF110 dataset~\cite{fraser2014evosuite}. However, the results from this dataset may not generalize when using other datasets. To investigate the generalizability, we additionally perform a study on the unseen Defects4J dataset~\cite{just2014defects4j}. 

%internal validity 
We developed several tools and scripts to automate various \mr{steps} and data analysis. While there is a possibility that our implementation may contain bugs, we took several precautions to minimize this risk. Each experiment was repeated at least three times to ensure result consistency. \mr{Furthermore, we utilized} publicly available libraries to reduce the likelihood of errors and performed comprehensive validity checks throughout the process.

\mrnew{Our qualitative evaluation involved interpreting \doc comments by the authors. To mitigate subjectivity, each author independently analyzed samples to identify helpful and harmful characteristics. We then integrated these findings to define the seven characteristics presented in RQ5 and used them for the larger qualitative analysis. While we did not analyze each sample by multiple authors, we cross-checked a random subset of each other's analyses to ensure consistency.}

%construct validity 
In our investigation, we utilized the \mr{\textit{exact match} metric during fine-tuning} to compute oracle generation accuracy, a widely \mr{adopted metric in} the research community. \mr{However, we acknowledge that this metric may underestimate accuracy, as correct test oracles can take many forms.} For unseen inference on the Defects4J dataset, we relied on test validation. \mr{We assert that this methodology offers a reliable accuracy metric, as it is sensitive \textit{only} to the correctness of the generated oracles, rather than their \textit{syntactic similarity}.}

%This methodology, we assert, \mr{provides} a reliable accuracy metric \mr{as} it is sensitive \mr{solely} to the \mr{correctness of the generated oracles, rather than their syntactical match.}

%\soneya{stopped here..........}
\section{Related Works} 

\mr{\subsection{Automated Test Oracle Generation}}
Automated test oracle generation (TOG) methods often employ a combination of approaches, broadly categorized into four types: fuzzing-based, search-based, specification mining-based, and machine learning-based. The fuzzing-based approach includes~\cite{pacheco2007randoop,zalewski2015afl}. Most fuzzing-based approaches rely on implicit oracles such as exceptions thrown due to unintended behavior (e.g., null dereferences, out-of-bounds exceptions). Search-based approach includes EvoSuite~\cite{fraser2011evosuite}, a SOTA method that generates regression oracles capable of detecting bugs in future versions of a program. However, this type of oracle cannot detect bugs in the current implementation, as it assumes that an executed behavior is the expected behavior. Specification mining-based methods leverage natural language processing and pattern recognition to derive test oracles from specifications~\cite{10.1145/3213846.3213872,10.1145/2931037.2931061,blasi2021memo,6227137,6200082,dinella2022toga}. A recent study has shown that these methods fail to generalize effectively and produce test oracles with subpar bug detection effectiveness~\cite{dinella2022toga}. Although the SOTA neural method TOGA was shown to significantly outperform existing neural~\cite{watson2020learning,tufano2022generating,tufano2020unit} and specification mining-based methods, and ~\cite{hossain2023neural} indicates that TOGA generates a high number of false positives and oracles that exhibit poor bug detection effectiveness. TOGLL~\cite{hossain2024togll} has shown to outperform TOGA by substantial margin.
\textit{Our goal in this paper is to dive deep into investigating the impact of \doc comment on the TOG task, rather than focusing strictly on outperforming other TOG methods -- though our method did substantially outperform TOGA on real-world fault detection}.

%Recently, \cite{endres2024can} conducted a study to investigate whether natural language documentation can be leveraged to generate test oracles. They utilized a small Python benchmark, EvalPlus, which includes 164 Python problems, each accompanied by a function stub and a natural language description in the form of a Python docstring. In their study, they employed OpenAI's GPT-3.5, GPT-4, and the open-source StarChat model to generate test oracles using few-shot prompting. In contrast, we utilize the SF110 dataset, a large-scale dataset of real-world projects, and instead of prompting, we fine-tune the models. Our experimental results demonstrate that fine-tuning is more effective than prompting in fully leveraging the power of LLMs.

%\mrnew{\subsection{Large Language Models for Software Engineering}
%Large Language Models (LLMs) have been applied to nearly all phases of software development lifecycle \cite{10.1145/3695988}, including specification generation \cite{xie2023impact}, code generation \cite{jiang2024ledex}, test generation \cite{dinella2022toga, hossain2024togll, siddiq2024using}, and bug repair \cite{10.1145/3660773}. While prompting LLMs using few-shot learning has been the predominant approach, recent studies have demonstrated the effectiveness of fine-tuning for teaching domain-specific patterns, leading to significant improvements in performance.}

\section{Conclusion}
In this work, we conduct a thorough investigation to understand the impact of \doc comments on the test oracle generation (TOG) task. Our large-scale experimental results demonstrate that high-quality \doc comments are valuable in automating test oracles, and a minimal prompt consisting solely of \doc comments can achieve performance comparable or even better than prompts that include maximum contextual information. Our analysis reveals that the ``description'' and \texttt{@return} tags are the most valuable for test oracle generation, and when there are constraints on prompt length, other details can be omitted without significantly impacting accuracy.   Furthermore, our study shows that \doc 
can be used to generate oracles that detect faults in real-world programs and, importantly, those oracles outperform oracles generated from MUT implementations.
This addresses a significant limitation of prior work, which our methods outperform.

\mrnew{This work is a first step in understanding the potential of \doc for TOG.
More work is needed to understand how to provide a rich, but minimal, \doc
context for specific TOG tasks.   For example, most Java
methods are written in the context of a class and relate to other classes.  Exploring
methods for identifying the dependencies among code elements, e.g., 
by following \texttt{@InheritDoc} or \texttt{@see} tags, and incorporating 
\doc for all of those elements has the potential to further boost TOG performance.}

\bibliographystyle{ACM-Reference-Format}
\bibliography{fse-25}

\end{document}

%% file: macros.tex
%--------------- TODO Commands --------------
\newcommand{\todoc}[2]{{\textcolor{#1}{#2}}}
\newcommand{\todoblack}[1]{{\todoc{black}{\textbf{[[#1]]}}}}
\newcommand{\todored}[1]{{\todoc{red}{\textbf{[[#1]]}}}}
\newcommand{\todogreen}[1]{\todoc{green}{\textbf{[[#1]]}}}
\newcommand{\todoblue}[1]{\todoc{black}{#1}}
\newcommand{\todoorange}[1]{\todoc{orange}{\textbf{[[#1]]}}}
\newcommand{\todobrown}[1]{\todoc{brown}{\textbf{[[#1]]}}}
\newcommand{\todogray}[1]{\todoc{gray}{\textbf{[[#1]]}}}
\newcommand{\todopurple}[1]{\todoc{purple}{\textbf{[[#1]]}}}
\newcommand{\todopink}[1]{\todoc{magenta}{\textbf{[[#1]]}}}
\newcommand{\todocyan}[1]{\todoc{cyan}{\textbf{[[#1]]}}}
\newcommand{\todoviolet}[1]{\todoc{violet}{\textbf{[[#1]]}}}
\newcommand{\todo}[1]{\todored{TODO: #1}}

\newcommand{\rt}[1]{\todopink{Raygan: #1}}
\newcommand{\soneya}[1]{\todogreen{Soneya: #1}}
\newcommand{\ignore}[1]{}
\newcommand{\raygan}[1]{\todoblue{#1}}
\newcommand{\toolname}{Toggle}
\newcommand{\matt}[1]{\todopurple{Matt: #1}}
\newcommand{\mr}[1]{{\color{black} #1}}
\newcommand{\mrnew}[1]{{\color{black} #1}}

\definecolor{col1}{RGB}{240, 240, 240} % Light Gray
\definecolor{col2}{RGB}{240, 240, 240} % Medium Gray
\definecolor{col3}{RGB}{220, 220, 220} % Darker Gray
\definecolor{col4}{RGB}{220, 220, 220} % Even Darker Gray
\definecolor{col5}{RGB}{200, 200, 200} % Very Light Gray
\definecolor{col6}{RGB}{200, 200, 200} % Medium Dark Gray

% ---------comment-----------

%% To disable colored comments, just uncomment this line: 
% \renewcommand{\todoc}[2]{\relax}

\newcommand{\code}[1]{\texttt{\small #1}} % code style

\newcounter{finding}
\newcommand{\finding}[1]{\refstepcounter{finding}
    \begin{mdframed}[linecolor=gray,roundcorner=12pt,backgroundcolor=gray!15,linewidth=3pt,innerleftmargin=2pt, leftmargin=0cm,rightmargin=0cm,topline=false,bottomline=false,rightline=false]
        \textbf{Finding \arabic{finding}:} #1
    \end{mdframed}
}